\documentclass[journal, letterpaper]{IEEEtran}
\usepackage[utf8]{inputenc}
\usepackage{blindtext}
\usepackage{algorithm2e}
\usepackage{graphicx}
\usepackage{amsmath}
\usepackage{amssymb}
\usepackage{hyperref}
\usepackage{csquotes}
\usepackage{enumerate}
\usepackage{siunitx}
\usepackage{accents}
\usepackage{subcaption}
\usepackage{algorithmic}
\usepackage{pgfplots}
\usepackage{amsthm}
\usepackage{todonotes}
\pgfplotsset{compat=1.14}
\usepackage{float}

\theoremstyle{definition}
\newtheorem{definition}{Definition}[]
\graphicspath{ {gfx/}{sketches/} }
\makeatletter
\def\input@path{{tikz/}}
\makeatother

\makeatletter \newcommand{\pgfplotsdrawaxis}{\pgfplots@draw@axis} \makeatother

\pgfplotsset{axis line on top/.style={
  axis line style=transparent,
  ticklabel style=transparent,
  tick style=transparent,
  axis on top=false,
  after end axis/.append code={
    \pgfplotsset{axis line style=opaque,
      ticklabel style=opaque,
      tick style=opaque,
      grid=none}
    \pgfplotsdrawaxis}
  }
}

\ifCLASSINFOpdf
\else
\fi
\begin{document}
\title{Cooperative Multi-Agent Reinforcement Learning for Low-Level Wireless Communication}

\author{Colin de Vrieze,
        Shane Barratt,
        Daniel Tsai and
        Anant Sahai~\IEEEmembership{UC Berkeley}}% <-this % stops a space
        
%\thanks{Manuscript received April 19, 2005; revised January 11, 2007.}}


\markboth{}%
{Shell \MakeLowercase{\textit{et al.}}: Bare Demo of IEEEtran.cls for Journals}



\maketitle

\begin{abstract}
%\boldmath

Traditional radio systems are strictly co-designed on the lower levels of the OSI stack for compatibility and efficiency. Although this has enabled the success of radio communications, it has also introduced lengthy standardization processes and imposed static allocation of the radio spectrum. Various initiatives have been undertaken by the research community to tackle the problem of artificial spectrum scarcity by both making frequency allocation more dynamic and building flexible radios to replace the static ones. There is reason to believe that just as computer vision and control have been overhauled by the introduction of machine learning, wireless communication can also be improved by utilizing similar techniques to increase the flexibility of wireless networks. In this work, we pose the problem of discovering low-level wireless communication schemes ex-nihilo between two agents in a fully decentralized fashion as a reinforcement learning problem. Our proposed approach uses policy gradients to learn an optimal bi-directional communication scheme and shows surprisingly sophisticated and intelligent learning behavior. We present the results of extensive experiments and an analysis of the fidelity of our approach. 
\end{abstract}

%\begin{IEEEkeywords}
%IEEEtran, journal, \LaTeX, paper, template.
%\end{IEEEkeywords}

\IEEEpeerreviewmaketitle

\section{Introduction}\label{sec:introduction}

In recent years, automatically learned features have replaced hand-designed filters in computer vision problems, yielding much higher accuracy on benchmark tasks such as image classification and object detection. These results are achieved by training a deep neural network end-to-end from raw pixels to output probabilities with a large amount of training data. In the process of optimizing the network to minimize prediction error, the model learns meaningful low-level spatial features in the form of activations in the hidden layers of the network~\cite{bengio2013representation}. There is no theoretical bound on performance, as more layers can always be added to increase the expressiveness of the model, and recent research using very deep networks has exceeded human level performance on tasks such as image classification \cite{inception2015}.

Data-driven reinforcement learning approaches have also proven to be an effective tool for approximately solving control problems such as object manipulation and robotic locomotion. Classically, these problems were solved by constructing the dynamics model of a system and optimizing the trajectory by using local linear models such as iLQR~\cite{bertsekas1995dynamic} or finite-difference methods. More modern approaches solve trajectory optimization by modelling the problem as an actor that inputs actions into the system and receives a new state and a reward signal as a response. The actor is then responsible for maximizing this reward signal. A derivative of this technique is multi-agent cooperative learning, which has recently been gaining in popularity as a method for solving simple logic games involving multiple actors~\cite{foerster}. 

Wireless communication is a domain that is traditionally characterized by manually designed signal processing blocks, similar to the hand-crafted features used in traditional computer vision. Low-level functions such as modulation and error correcting codes have been carefully designed to optimize the performance of the radios under various channel conditions. 

\begin{table}[!t]
\fontsize{9}{9}\selectfont
\renewcommand{\arraystretch}{1.3}
\centering
\begin{tabular}{c c c}
\hline
\textbf{DSP tool} & & \textbf{ML equivalent} \\
\hline\hline
FIR filter & $\longleftrightarrow$ & Convolutional layer \\
IIR filter & $\longleftrightarrow$ & Recurrent cell \\
LMS equalizer & $\longleftrightarrow$ & Gradient descent algorithm \\
Source coding & $\longleftrightarrow$ & Autoencoders
\end{tabular}
\caption{Parallels between classic digital signal processing (DSP) and modern machine learning (ML) tools. A FIR filter convolves a signal with a kernel, just as a convolutional layer in a neural network does. An IIR Filter is a filter with feedback, just as a recurrent cell performs computation with feedback. The least mean squares (LMS) equalizer updates the parameters of an equalization filter using gradient descent, just as gradient descent is used to optimize the parameters of neural networks. Finally, source coding (compression) seeks to find a "smaller" representation of data, just as an autoencoder attempts to compress input data through a bottleneck and then decode it.}
\label{tab:dspml}
\end{table}

In addition, traditional radio systems are strictly co-designed on the lower levels of the OSI stack for compatibility and efficiency. Although this has enabled radio communications to succeed despite hardware constraints, it has also introduced lengthy standardization processes and imposed static allocation of the radio spectrum. Various initiatives have been undertaken by the research community to tackle the problem of artificial spectrum scarcity by making frequency allocation more dynamic and building collaborative networks to replace the static ones. The most prominent of these initiatives is the DARPA Spectrum Collaboration Challenge (SC2), which is \enquote{the first-of-its-kind collaborative machine-learning competition to overcome scarcity in the radio frequency (RF) spectrum}.\footnote{\href{www.spectrumcollaborationchallenge.com}{spectrumcollaborationchallenge.com}}

In this work, we investigate the use of modern reinforcement learning techniques to alleviate the rigidity of traditional radio network design. We intend to replace ordinary radio blocks such as modulation and demodulation with high-capacity models that are agnostic to their specific function, and are instead learned in a data-driven manner. The parallels between classic digital signal processing blocks and machine learning tools (see Table~\ref{tab:dspml}) justify the assumption that high-capacity models should be able to learn low-level radio functions. In contrast to other domains, data for training can be easily generated for large-scale simulations. Finally, we will show that it is possible for two actors to learn modulation schemes for communication while sharing only a fixed bit string and having no domain-specific knowledge about the task.

The rest of the work is organized as follows: Section~\ref{sec:related-work} and Section~\ref{sec:background} discuss related work and give some background information on wireless communication and reinforcement learning. Section~\ref{sec:preliminary analysis} presents some preliminary analysis of data driven approaches to classic radio tasks. Section~\ref{sec:main-problem-formulation} introduces the problem of a fully decentralized multi-agent setting for learning modulation. Section~\ref{sec:main-problem-setup} and~\ref{sec:results} present our solution and results, respectively. Finally, some concluding remarks are given in Section~\ref{sec:conclusion}.

\section{Related Work}\label{sec:related-work}

The research presented in this work spans many fields, so the related works will be grouped by research topic. First, we review early approaches to parameter optimization in radio-related tasks, then we explore machine learning algorithms in the communication domain and finish by describing recent work on distributed deep reinforcement learning.

The concept of using parameter optimization in radio-related tasks was first introduced in 1960~\cite{widrow}. Widrow applied gradient descent to the problem of equalization in the form of the least mean squares equalizer, which optimizes the filter parameters of a fixed length FIR filter to counteract the effects of a multipath channel (see Section~\ref{sec:commsbg}). His work also links back to the optimization of single perceptrons. In 1957, Lloyd came up with his algorithm for k-means unsupervised learning while solving the problem of demodulating pulse-code modulation. His work was published in 1982~\cite{kmeans} and now serves as the standard solution to solving k-means. Through their interdisciplinary work, Widrow and Lloyd were pioneers in both digital wireless communication and machine learning.

Since Widrow's early work, research into the use of neural architectures and machine learning techniques to address problems in communications has taken off in many directions. Ibnkahla's comprehensive review of the intersection of communications and machine learning~\cite{ibnkahla} is a testament to the impressive efforts made in this area of research. His survey lists various learning-based approaches to adaptive equalization, nonlinear channel modeling, coding, error correcting codes, spread spectrum applications, network planning, modulation detection and many more. Due to the vastness of this field, we refer the reader to the review. Research in making radio agents adaptive to achieve cooperative goals began in 1999, when Mitola introduced the concept of the cognitive radio~\cite{mitola1999cognitive}. His proposal was for cognitive radios to use model-based reasoning to achieve competency in radio related tasks using both supervised and unsupervised learning. A review of cognitive radio work in years after that is in~\cite{zeng2010review}.

Beyond the applications to game-playing and control tasks, some researchers have begun to investigate the use of reinforcement learning in various communication-based cooperative multi-agent tasks. Foerster et al.~\cite{foerster} applied variants of deep Q-networks (DQN) without experience replay to prisoner's games with multiple agents. The considered problems require the agents to communicate over a very simple noiseless channel of small bandwidth and develop a collaborative strategy. Mordatch and Abbeel~\cite{openai} show how a grounded, compositional language can emerge when agents have to communicate their intentions to each other in order to maximize their reward. Finally, Abadi and Andersen~\cite{crypto} study the problem of learning encryption between agents with the help of an adversary. 

Though these works are not concerned with the low level details of wirelessly transmitting digital signals, they do incorporate communication between the agents as an essential part to solve the presented problems. In all of them, the communication protocol is not given upfront but has to be figured out during training. However, their works rely on several assumptions which set them apart from a fully decentralized setting: Foerster et al. implement decentralized execution but centralized training by using a single set of parameters for all agents. Mordatch and Abbeel's work also trains a single policy, but assumes a fully differentiable environment and communication protocol. Lastly, Abadi and Andersen relax the problem by allowing agents to exchange rewards and parameters. In contrast, our approach aims to fully decentralize the learning process to minimize the gap to a real system. It gives hope that fully distributed deep reinforcement learning is possible.

\section{Background}\label{sec:background}
\subsection{Low Level Digital Wireless Communication}\label{sec:commsbg}

\begin{figure}[!ht]
\centering
\includegraphics[width=3.5in]{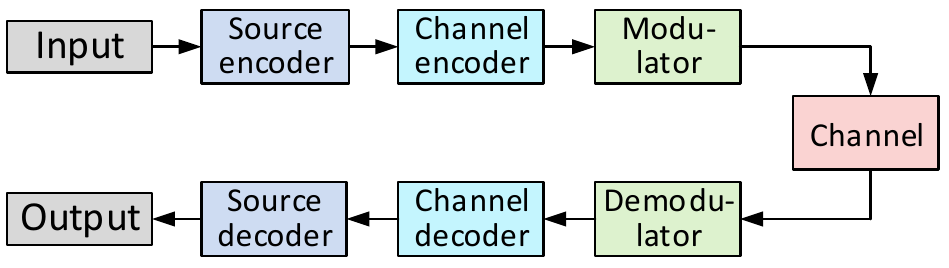}
  \caption{Simple model of a communication link between a transmitter (top) and a receiver (bottom). In the transmitter, source coding removes redundancies from the source signal  while channel coding systematically adds redundancy for error correction. Modulation is used to impose the information on the carrier, which conveys it through the channel. On the receiver side, all processes are reversed to reconstruct the original signal.}
  \label{fig:chain}
\end{figure}

Figure~\ref{fig:chain} shows a simplified model of a generic communication link. The goal in every setting is to transmit information from some source (input) across a \textit{channel} to some sink (output) with the help of a \textit{carrier}. The channel is an abstract description of a means to transmit the information between the transmitter (upper blocks) and the receiver (lower blocks). It typically corrupts the signal in some way, e.g. by adding delays, noise or echos, which stem from the physical nature of the carrier.  

\begin{table}[!ht]
\fontsize{9}{9}\selectfont
\renewcommand{\arraystretch}{1.3}
\centering
\begin{tabular}{c c c c}
\hline
\textbf{Input} & \textbf{Channel} & \textbf{Carrier} & \textbf{Output} \\
\hline\hline
Microphone & Hard drive & Magnetic remanence & Speaker \\
Camera & Copper cable & Voltage signal & Screen \\
Radio & Wireless link & Electromagnetic wave & Radio \\
\end{tabular}%
\caption{Examples of typical communication systems. We focus on the setting with two radios and a wireless link.}%
\label{tab:channels}%
\end{table}

In this work, we ignore the problem of source and channel coding and focus on the modulation and demodulation of the the carrier. Wireless communication utilizes electromagnetic waves as a carrier to convey information across a wireless link between radios. Those waves are described by sinusoidal functions in time and space. Ignoring the details of propagation, the wave induces a voltage signal $s(t)$ in an antenna at a given point in space. We can interpret the signal $s(t)$ as a sum of two parts: The \textit{in-phase} component $I(t)$ and the \textit{quadrature} component $Q(t)$ plus the sinusoidal functions $cos(2\pi ft)$ and $sin(2\pi ft)$, respectively. Because $sin$ and $cos$ are orthogonal over multiples of a period $T=1/f$, the receiver of the electromagnetic wave can extract the two functions $I(t)$ and $Q(t)$ from the signal $s(t)$ independently. Thus, we can interpret the components as two independent degrees of freedom to use for the transmission of information.

\begin{equation} \label{eq:signal}
\begin{aligned}
s(t) & = I(t)cos(2\pi ft)-Q(t)sin(2\pi ft) \\
     & = Re\{[I(t)+iQ(t)]e^{i2\pi ft}\}
\end{aligned}
\end{equation}

\begin{figure}[h]
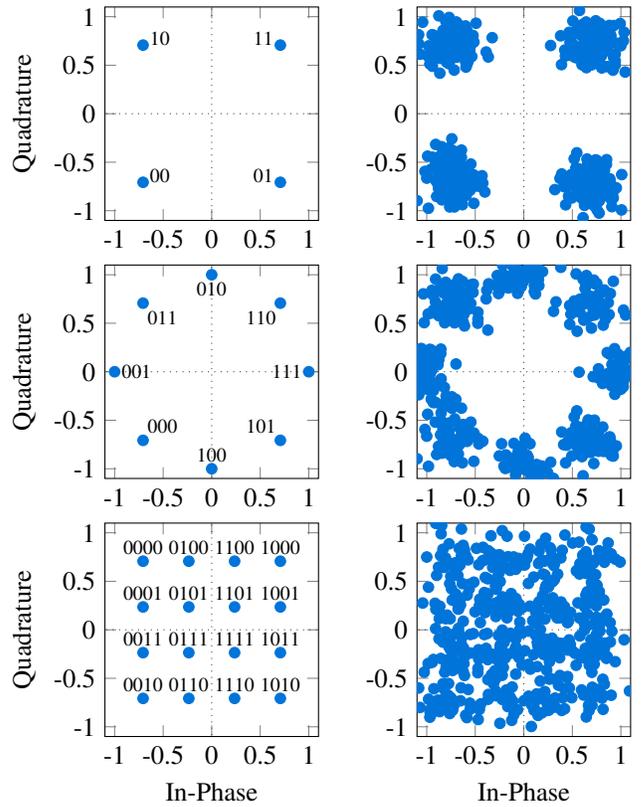

\begin{subfigure}{0.24\textwidth}
\centering
% This file was created by matplotlib2tikz v0.6.6.
\begin{tikzpicture}

\definecolor{color0}{rgb}{0,0.454901960784314,0.850980392156863}
\definecolor{grey}{rgb}{0.25,0.25,0.25}

\begin{axis}[
ylabel={Quadrature},
xmin=-1.1, xmax=1.1,
ymin=-1.1, ymax=1.1,
%axis on top,
width=\textwidth,
height=\textwidth,
xtick={-1.5,-1,-0.5,0,0.5,1,1.5},
xticklabels={-1.5,-1,-0.5,0,0.5,1,1.5},
ytick={-1.5,-1,-0.5,0,0.5,1,1.5},
yticklabels={-1.5,-1,-0.5,0,0.5,1,1.5},
tick pos=both
]

\addplot[dotted, color=grey] coordinates {(-2,0) (2,0)};
\addplot[dotted, color=grey] coordinates {(0,-2) (0,2)};

\addplot [only marks, draw=color0, fill=color0]
table {%
x                      y
-7.071067811865475e-01 -7.071067811865475e-01
+7.071067811865475e-01 -7.071067811865475e-01
-7.071067811865475e-01 +7.071067811865475e-01
+7.071067811865475e-01 +7.071067811865475e-01
};

\draw[] (axis cs:-0.707106781186547,-0.707106781186547) -- (axis cs:-0.707106781186547,-0.707106781186547);
\node at (axis cs:-0.707106781186547,-0.707106781186547)[
  scale=0.75,
  anchor=base west,
  text=black,
  rotate=0.0
]{00};
\draw[] (axis cs:0.707106781186547,-0.707106781186547) -- (axis cs:0.707106781186547,-0.707106781186547);
\node at (axis cs:0.707106781186547,-0.707106781186547)[
  scale=0.75,
  anchor=base east,
  text=black,
  rotate=0.0
]{01};
\draw[] (axis cs:-0.707106781186547,0.707106781186547) -- (axis cs:-0.707106781186547,0.707106781186547);
\node at (axis cs:-0.707106781186547,0.707106781186547)[
  scale=0.75,
  anchor=base west,
  text=black,
  rotate=0.0
]{10};
\draw[] (axis cs:0.707106781186547,0.707106781186547) -- (axis cs:0.707106781186547,0.707106781186547);
\node at (axis cs:0.707106781186547,0.707106781186547)[
  scale=0.75,
  anchor=base east,
  text=black,
  rotate=0.0
]{11};
\end{axis}

\end{tikzpicture}
\end{subfigure}%
\begin{subfigure}{0.24\textwidth}
\centering
\input{qpsk_noise.tex}
\end{subfigure}%

\begin{subfigure}{0.24\textwidth}
\centering
% This file was created by matplotlib2tikz v0.6.6.
\begin{tikzpicture}

\definecolor{color0}{rgb}{0,0.454901960784314,0.850980392156863}
\definecolor{grey}{rgb}{0.25,0.25,0.25}

\begin{axis}[
ylabel={Quadrature},
xmin=-1.1, xmax=1.1,
ymin=-1.1, ymax=1.1,
%axis on top,
width=\textwidth,
height=\textwidth,
xtick={-1.5,-1,-0.5,0,0.5,1,1.5},
xticklabels={-1.5,-1,-0.5,0,0.5,1,1.5},
ytick={-1.5,-1,-0.5,0,0.5,1,1.5},
yticklabels={-1.5,-1,-0.5,0,0.5,1,1.5},
tick pos=both
]

\addplot[dotted, color=grey] coordinates {(-2,0) (2,0)};
\addplot[dotted, color=grey] coordinates {(0,-2) (0,2)};

\addplot [only marks, draw=color0, fill=color0]
table {%
x                      y
-7.071067811865475e-01 -7.071067811865475e-01
-1.000000000000000e+00 +0.000000000000000e+00
+0.000000000000000e+00 +1.000000000000000e+00
-7.071067811865475e-01 +7.071067811865475e-01
+0.000000000000000e+00 -1.000000000000000e+00
+7.071067811865475e-01 -7.071067811865475e-01
+7.071067811865475e-01 +7.071067811865475e-01
+1.000000000000000e+00 +0.000000000000000e+00
};

\draw[] (axis cs:-0.707106781186547,-0.707106781186547) -- (axis cs:-0.707106781186547,-0.707106781186547);
\node at (axis cs:-0.507106781186547,-0.707106781186547)[
  scale=0.75,
  anchor=south,
  text=black,
  rotate=0.0
]{000};
\draw[] (axis cs:-1,0) -- (axis cs:-1,0);
\node at (axis cs:-1,0)[
  scale=0.75,
  anchor=west,
  text=black,
  rotate=0.0
]{001};
\draw[] (axis cs:0,1) -- (axis cs:0,1);
\node at (axis cs:0,1)[
  scale=0.75,
  anchor=north,
  text=black,
  rotate=0.0
]{010};
\draw[] (axis cs:-0.707106781186547,0.707106781186547) -- (axis cs:-0.707106781186547,0.707106781186547);
\node at (axis cs:-0.507106781186547,0.697106781186547)[
  scale=0.75,
  anchor=north,
  text=black,
  rotate=0.0
]{011};
\draw[] (axis cs:0,-1) -- (axis cs:0,-1);
\node at (axis cs:0,-1)[
  scale=0.75,
  anchor=south,
  text=black,
  rotate=0.0
]{100};
\draw[] (axis cs:0.707106781186547,-0.707106781186547) -- (axis cs:0.707106781186547,-0.707106781186547);
\node at (axis cs:0.507106781186547,-0.707106781186547)[
  scale=0.75,
  anchor=south,
  text=black,
  rotate=0.0
]{101};
\draw[] (axis cs:0.707106781186547,0.707106781186547) -- (axis cs:0.707106781186547,0.707106781186547);
\node at (axis cs:0.507106781186547,0.697106781186547)[
  scale=0.75,
  anchor=north,
  text=black,
  rotate=0.0
]{110};
\draw[] (axis cs:1,0) -- (axis cs:1,0);
\node at (axis cs:1,0)[
  scale=0.75,
  anchor=east,
  text=black,
  rotate=0.0
]{111};
\end{axis}

\end{tikzpicture}
\end{subfigure}%
\begin{subfigure}{0.24\textwidth}
\centering
\input{8psk_noise.tex}
\end{subfigure}%

\begin{subfigure}{0.24\textwidth}
\centering
% This file was created by matplotlib2tikz v0.6.6.
\begin{tikzpicture}

\definecolor{color1}{rgb}{1,0.254901960784314,0.211764705882353}
\definecolor{color0}{rgb}{0,0.454901960784314,0.850980392156863}
\definecolor{grey}{rgb}{0.25,0.25,0.25}

\begin{axis}[
xlabel={In-Phase},
ylabel={Quadrature},
xmin=-1.1, xmax=1.1,
ymin=-1.1, ymax=1.1,
%axis on top,
width=\textwidth,
height=\textwidth,
xtick={-1.5,-1,-0.5,0,0.5,1,1.5},
xticklabels={-1.5,-1,-0.5,0,0.5,1,1.5},
ytick={-1.5,-1,-0.5,0,0.5,1,1.5},
yticklabels={-1.5,-1,-0.5,0,0.5,1,1.5},
tick pos=both
]

\addplot[dotted, color=grey] coordinates {(-2,0) (2,0)};
\addplot[dotted, color=grey] coordinates {(0,-2) (0,2)};

\addplot [only marks, draw=color0, fill=color0, colormap/viridis]
table {%
x                      y
-7.071067811865475e-01 +7.071067811865475e-01
-7.071067811865475e-01 +2.357022603955158e-01
-7.071067811865475e-01 -7.071067811865475e-01
-7.071067811865475e-01 -2.357022603955158e-01
-2.357022603955158e-01 +7.071067811865475e-01
-2.357022603955158e-01 +2.357022603955158e-01
-2.357022603955158e-01 -7.071067811865475e-01
-2.357022603955158e-01 -2.357022603955158e-01
+7.071067811865475e-01 +7.071067811865475e-01
+7.071067811865475e-01 +2.357022603955158e-01
+7.071067811865475e-01 -7.071067811865475e-01
+7.071067811865475e-01 -2.357022603955158e-01
+2.357022603955158e-01 +7.071067811865475e-01
+2.357022603955158e-01 +2.357022603955158e-01
+2.357022603955158e-01 -7.071067811865475e-01
+2.357022603955158e-01 -2.357022603955158e-01
};

\draw[] (axis cs:-0.707106781186547,0.707106781186547) -- (axis cs:-0.707106781186547,0.707106781186547);
\node at (axis cs:-0.707106781186547,0.707106781186547)[
  scale=0.75,
  anchor=south,
  text=black,
  rotate=0.0
]{0000};
\draw[] (axis cs:-0.707106781186547,0.235702260395516) -- (axis cs:-0.707106781186547,0.235702260395516);
\node at (axis cs:-0.707106781186547,0.235702260395516)[
  scale=0.75,
  anchor=south,
  text=black,
  rotate=0.0
]{0001};
\draw[] (axis cs:-0.707106781186547,-0.707106781186547) -- (axis cs:-0.707106781186547,-0.707106781186547);
\node at (axis cs:-0.707106781186547,-0.707106781186547)[
  scale=0.75,
  anchor=south,
  text=black,
  rotate=0.0
]{0010};
\draw[] (axis cs:-0.707106781186547,-0.235702260395516) -- (axis cs:-0.707106781186547,-0.235702260395516);
\node at (axis cs:-0.707106781186547,-0.235702260395516)[
  scale=0.75,
  anchor=south,
  text=black,
  rotate=0.0
]{0011};
\draw[] (axis cs:-0.235702260395516,0.707106781186547) -- (axis cs:-0.235702260395516,0.707106781186547);
\node at (axis cs:-0.235702260395516,0.707106781186547)[
  scale=0.75,
  anchor=south,
  text=black,
  rotate=0.0
]{0100};
\draw[] (axis cs:-0.235702260395516,0.235702260395516) -- (axis cs:-0.235702260395516,0.235702260395516);
\node at (axis cs:-0.235702260395516,0.235702260395516)[
  scale=0.75,
  anchor=south,
  text=black,
  rotate=0.0
]{0101};
\draw[] (axis cs:-0.235702260395516,-0.707106781186547) -- (axis cs:-0.235702260395516,-0.707106781186547);
\node at (axis cs:-0.235702260395516,-0.707106781186547)[
  scale=0.75,
  anchor=south,
  text=black,
  rotate=0.0
]{0110};
\draw[] (axis cs:-0.235702260395516,-0.235702260395516) -- (axis cs:-0.235702260395516,-0.235702260395516);
\node at (axis cs:-0.235702260395516,-0.235702260395516)[
  scale=0.75,
  anchor=south,
  text=black,
  rotate=0.0
]{0111};
\draw[] (axis cs:0.707106781186547,0.707106781186547) -- (axis cs:0.707106781186547,0.707106781186547);
\node at (axis cs:0.707106781186547,0.707106781186547)[
  scale=0.75,
  anchor=south,
  text=black,
  rotate=0.0
]{1000};
\draw[] (axis cs:0.707106781186547,0.235702260395516) -- (axis cs:0.707106781186547,0.235702260395516);
\node at (axis cs:0.707106781186547,0.235702260395516)[
  scale=0.75,
  anchor=south,
  text=black,
  rotate=0.0
]{1001};
\draw[] (axis cs:0.707106781186547,-0.707106781186547) -- (axis cs:0.707106781186547,-0.707106781186547);
\node at (axis cs:0.707106781186547,-0.707106781186547)[
  scale=0.75,
  anchor=south,
  text=black,
  rotate=0.0
]{1010};
\draw[] (axis cs:0.707106781186547,-0.235702260395516) -- (axis cs:0.707106781186547,-0.235702260395516);
\node at (axis cs:0.707106781186547,-0.235702260395516)[
  scale=0.75,
  anchor=south,
  text=black,
  rotate=0.0
]{1011};
\draw[] (axis cs:0.235702260395516,0.707106781186547) -- (axis cs:0.235702260395516,0.707106781186547);
\node at (axis cs:0.235702260395516,0.707106781186547)[
  scale=0.75,
  anchor=south,
  text=black,
  rotate=0.0
]{1100};
\draw[] (axis cs:0.235702260395516,0.235702260395516) -- (axis cs:0.235702260395516,0.235702260395516);
\node at (axis cs:0.235702260395516,0.235702260395516)[
  scale=0.75,
  anchor=south,
  text=black,
  rotate=0.0
]{1101};
\draw[] (axis cs:0.235702260395516,-0.707106781186547) -- (axis cs:0.235702260395516,-0.707106781186547);
\node at (axis cs:0.235702260395516,-0.707106781186547)[
  scale=0.75,
  anchor=south,
  text=black,
  rotate=0.0
]{1110};
\draw[] (axis cs:0.235702260395516,-0.235702260395516) -- (axis cs:0.235702260395516,-0.235702260395516);
\node at (axis cs:0.235702260395516,-0.235702260395516)[
  scale=0.75,
  anchor=south,
  text=black,
  rotate=0.0
]{1111};
\end{axis}

\end{tikzpicture}
\end{subfigure}%
\begin{subfigure}{0.24\textwidth}
\centering
\input{16qam_noise.tex}
\end{subfigure}%

\caption{Left column: Common modulation schemes: QPSK (2 bits per symbol), 8-PSK (3 bits per symbol) and 16-QAM (4 bits per symbol). Right column: Impact of AWGN with noise power density $N_0=0.04$ on the received signal.}
\label{fig:modulations}

\end{figure}

For the purpose of dealing with the in-phase and quadrature component, Equation~\ref{eq:signal} shows how to think of the signal $s(t)$ as the real part of a complex number $x(t)=I(t)+iQ(t)$ multiplied by a carrier signal $c(t)=e^{i2\pi ft}$. We say the signal $x(t)$ \textit{modulates} the carrier $c(t)$. In digital modulation, we allow only certain values for $x(t)$. The plot of all valid values for $x(t)$ in the complex plane is called a \textit{constellation diagram} and each individual value results in a \textit{constellation point}. Given the maximum number of valid points for a certain \textit{modulation scheme}, we can assign a bit sequence with a set length to each of the points individually. For transmitting information, the transmitter modulates the electromagnetic wave by setting the value of $x(t)$ to a constellation point for a certain duration $T_s$ of one \textit{symbol}. A series of bits can be transmitted by consecutively sending symbols, each of which conveys a given number of bits. The transmitter and receiver have to agree upfront on the modulation scheme as well as the mapping between constellation points and bit sequences to successfully transmit a message. Figure~\ref{fig:modulations} shows common modulation schemes for up to four bits per symbol. 

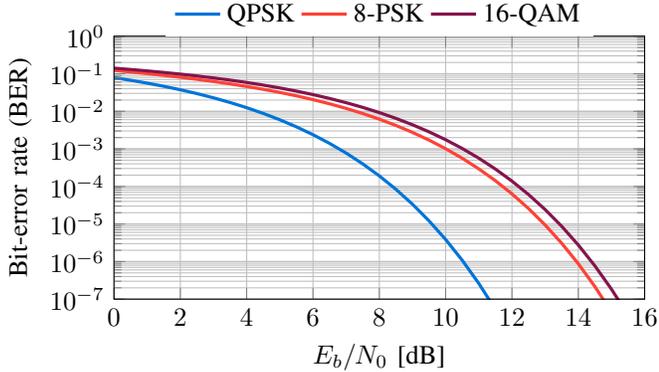
\begin{figure}[!ht]
\centering
% This file was created by matplotlib2tikz v0.6.6.
\begin{tikzpicture}

\definecolor{red}{rgb}{1,0.254901960784314,0.211764705882353}
\definecolor{blue}{rgb}{0,0.454901960784314,0.850980392156863}
\definecolor{purple}{rgb}{0.52156862745098,0.0784313725490196,0.294117647058824}

\begin{axis}[
xlabel={$E_b/N_0$ [dB]},
ylabel={Bit-error rate (BER)},
xmin=0, xmax=16,
ymin=1e-07, ymax=1,
ymode=log,
axis line on top,
grid=both,
width=3.4in,
height=2in,
ytick={1e-08,1e-07,1e-06,1e-05,0.0001,0.001,0.01,0.1,1,10},
yticklabels={0.0,$ 10^{-7}$,$ 10^{-6}$,$ 10^{-5}$,$ 10^{-4}$,$ 10^{-3}$,$ 10^{-2}$,$ 10^{-1}$,$ 10^{0}$,},
tick pos=both,
legend entries={{QPSK},{8-PSK},{16-QAM}},
legend style={at={(0.5,1)},
anchor=south,
draw=none,
%text height=0.7ex
},
legend columns=3,
legend cell align={left}
]
\addplot [very thick, blue, on layer=main]
table {%
0	0.0786496035251426
0.500000000000000	0.0670651983296128
1	0.0562819519765415
1.50000000000000	0.0464012759560713
2	0.0375061283589260
2.50000000000000	0.0296552876260368
3	0.0228784075610853
3.50000000000000	0.0171725416792460
4	0.0125008180407376
4.50000000000000	0.00879381053056083
5	0.00595386714777866
5.50000000000000	0.00386223164281013
6	0.00238829078093281
6.50000000000000	0.00139980483948023
7	0.000772674815378444
7.50000000000000	0.000398796335159163
8	0.000190907774075993
8.50000000000000	8.39995391790110e-05
9	3.36272284196176e-05
9.50000000000000	1.21088932770337e-05
10	3.87210821552205e-06
10.5000000000000	1.08384842174308e-06
11	2.61306795357521e-07
11.5000000000000	5.32866403631074e-08
12	9.00601035062878e-09
12.5000000000000	1.23302844661081e-09
13	1.33293101753005e-10
13.5000000000000	1.10545978569829e-11
14	6.81018912878076e-13
14.5000000000000	3.00554978904599e-14
15	9.12395736262818e-16
15.5000000000000	1.82023424951961e-17
16	2.26739584445444e-19
};
\addplot [very thick, red]
table {%
0	0.122692761078506
0.500000000000000	0.111540652877562
1	0.100798515869953
1.50000000000000	0.0904835695617223
2	0.0806094135504400
2.50000000000000	0.0711903834129154
3	0.0622456449843316
3.50000000000000	0.0538020474675053
4	0.0458949184655364
4.50000000000000	0.0385663949576311
5	0.0318614414208807
5.50000000000000	0.0258222386025809
6	0.0204819662829132
6.50000000000000	0.0158590671506789
7	0.0119529022708210
7.50000000000000	0.00874140828934567
8	0.00618105608376764
8.50000000000000	0.00420913573716912
9	0.00274813358917446
9.50000000000000	0.00171169468318278
10	0.00101139532098867
10.5000000000000	0.000563359097814931
11	0.000293729252229170
11.5000000000000	0.000142218949862518
12	6.33787882330033e-05
12.5000000000000	2.57371007802822e-05
13	9.41726453026406e-06
13.5000000000000	3.06592638195031e-06
14	8.75632695831043e-07
14.5000000000000	2.15925668121585e-07
15	4.51609253538730e-08
15.5000000000000	7.85243371235657e-09
16	1.10986997823151e-09
};

\addplot [very thick, purple]
table {%
0	0.140981635066842
0.500000000000000	0.129902828985330
1	0.118997407465924
1.50000000000000	0.108270144664669
2	0.0977418537374870
2.50000000000000	0.0874510119613148
3	0.0774530602925490
3.50000000000000	0.0678175657106818
4	0.0586237372834044
4.50000000000000	0.0499549326059661
5	0.0418927600464623
5.50000000000000	0.0345112383004777
6	0.0278713278451503
6.50000000000000	0.0220160860231097
7	0.0169667343687604
7.50000000000000	0.0127200094025227
8	0.00924721374147442
8.50000000000000	0.00649532590125164
9	0.00439033608735211
9.50000000000000	0.00284266538383420
10	0.00175415061789273
10.5000000000000	0.00102572522794619
11	0.000564706106481744
11.5000000000000	0.000290604575524983
12	0.000138658688812619
12.5000000000000	6.07855540662083e-05
13	2.42337854663159e-05
13.5000000000000	8.68609085076228e-06
14	2.76320800168778e-06
14.5000000000000	7.68966535739691e-07
15	1.84185551109448e-07
15.5000000000000	3.72859128323722e-08
16	6.25020082774196e-09
};
\end{axis}

\end{tikzpicture}
  \caption{Performance comparison of QPSK, 8-PSK and 16-QAM in terms of BER for different noise levels.}
  %A high $E_b/N_0$ ratio means relatively low noise and facilitates a low BER.}%
  \vspace{-.2cm}
  \label{fig:ber_comp}%
\end{figure}

The radio hardware as well as the signal propagation through the wireless channel are not perfect but instead attenuate and corrupt the signal. Various  effects can be simplified to an additive white Gaussian noise (AWGN) process with zero mean and variance $\sigma^2=N_0/2$ on both in-phase and quadrature component. On the receiver side, the noise leads to bit errors if a symbol is accidentally mapped to the wrong constellation point due to the noise. In the normalized constellation diagram (see Figure~\ref{fig:modulations}), the constellation points of, e.g. QPSK, are further apart than those of 16-QAM. Hence, for the same noise power density\footnote{\enquote{Noise power density} refers to the power spectral density of the noise and is the given noise power per unit of bandwidth. To calculate the total noise power one has to integrate over the bandwidth of the signal.} $N_0$, misclassifications of symbols are less likely for QPSK than for 16-QAM modulation. However, 16-QAM conveys twice the amount of bits per symbol and consequentially doubles the bit rate. Thus, there obviously exists a trade-off between the bit rate and resilience against bit errors among the different modulation schemes. Higher noise usually necessitates the use of a lower order modulation and hence lower bit rate. To improve the resilience of the transmission, it furthermore makes sense to map bit sequences with minimal Hamming distance\footnote{The Hamming distance $d$ of two bit sequences is defined as the number of dissimilar bits between them. For example: $d(0011,1010) = 2$.} to adjacent constellation points. In that way, the number of bit errors is minimized if a certain symbol gets confused with its neighbor. A mapping scheme which guarantees that every pair of directly neighboring constellation points has a Hamming distance of $d_{min}=1$ is called Gray coding.
\newpage
To quantify the performance of different modulation schemes under noise one usually examines the mean bit error rate (BER) across different noise situations given by the $E_b/N_0$ ratio. $E_b$ is the mean energy per bit and $N_0$ is the noise power density. $E_b/N_0$ is a normalized signal-to-noise (SNR) ratio, also known as the \enquote{SNR-per-bit}. Figure~\ref{fig:ber_comp} shows a performance comparison of traditional modulations schemes across different noise levels.

\begin{figure}[!ht]
\centering
\includegraphics[width=3.5in]{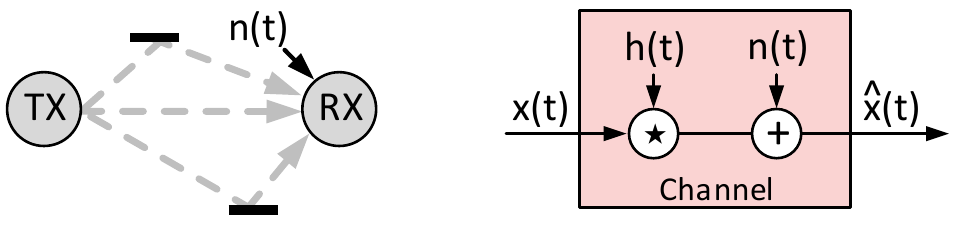}
  \caption{Left: Multipath propagation of the wireless signal leads to a superposition of the signal and delayed and attenuated copies of itself at the receiver. Right: $\hat{x}(t)$ can be modeled by a convolution of $x(t)$ with $h(t)$ and the addition of a complex Gaussian random variable $n(t)\sim\mathcal{C}\mathcal{N}(0,N_0)$.}
  \label{fig:channel}
    \vspace{-.2cm}
\end{figure}

In wireless communication, the signal does not only propagate directly from transmitter to receiver through the line-of-sight path, but it also gets reflected from objects in the environment (see Figure~\ref{fig:channel}). Thus, at every time step, the receiver receives a superposition of the signal and delayed and attenuated copies plus additive noise, which leads to inter-symbol interference. The effect of multipath propagation is equivalent to a convolution of the input signal with the \textit{channel impulse response} $h(t)$, which characterizes the wireless environment. Thus, the output signal of the channel can be modeled as $\hat{x}(t)=x(t)*h(t)+n(t)$. 

\begin{figure}[!ht]
\centering
\includegraphics[width=2.2in]{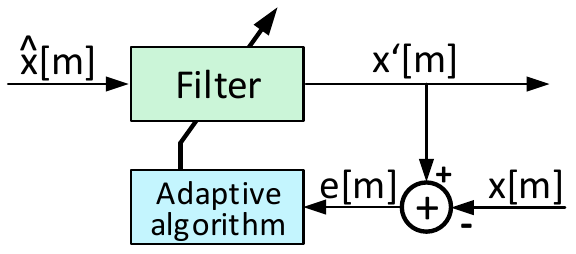}
  \caption{Principle of the LMS equalizer for discrete-time samples $x[m]$ of the continuous signal $x(t)$. Initialized with weights $w=w_1,...,w_N$, the filter outputs the convoluted signal $x'[m]=\hat{x}[m]*w[m]$ given the corrupted signal $\hat{x}[m]$. During training, the correct signal $x[m]$ is known, thus the equalizer can calculate the error between the reconstructed signal and the correct signal: $e[m]=x'[m]-x[m]$. The adaptive algorithm then updates the weights of the filter with learning rate $\alpha$ in the direction of steepest descent of the energy of the error signal: $w_{t+1}\leftarrow w_t-\alpha \frac{d(|e|)^2}{dw}$.}
  %$\alpha$ has to be tuned to enable fast adaptation but prevent divergence.}
  \label{fig:lms}
\vspace{-.2cm}
\end{figure}

If uncompensated, the inter-symbol interference would lead to detection errors, so it has to be dealt with before demodulation, which is called \textit{equalization}. Figure~\ref{fig:lms} shows one of the first approaches to equalization: The least mean squared~(LMS) equalizer developed by Widrow and Hoff in 1960~\cite{widrow}. Interestingly, both the LMS equalizer and neural architectures use gradient descent to update their parameters. Links to both of these applications can be found in Widrow's seminal work. This fact once again raises the question of how well modern machine learning approaches perform in digital radio tasks since both share a common ancestor.

\subsection{Reinforcement Learning and Policy Gradients}\label{subsec:rlpolicy}

One of the most difficult problems in the field of artificial intelligence today is that of sequential decision making in stochastic systems. These problems are characterized by an environment whose state evolves as a probabilistic function of the actions taken by the actor. Reinforcement learning algorithms direct an agent to choose optimal actions. In this work, we pose the problem of learning low-level wireless communication schemes in a decentralized fashion as a reinforcement learning problem and propose the policy gradient algorithm as a solution. This section introduces Markov decision processes, policy search, the score function gradient estimator and a vanilla policy gradient algorithm as they are all central to our approach.

Markov decision processes (MDPs) are a formalism for reasoning about decision making under uncertainty. There is an environment that takes on a set of states and an actor that takes actions at discrete time steps. Based on the state and the action taken, every time step, the actor receives a scalar reward. MDPs satisfy the Markov property, that all information relevant to the dynamics of the system define the state and only the current state and action affect the transition to the next state. An MDP can be formally defined as tuple $(S, D, A, {P(s'|s, a)}, P(R|s, a))$:

\begin{itemize}
\item $S$: A set of possible states of the world.
\item $D$: A probability distribution over the set of initial states.
\item $A$: The set of possible actions.
\item ${P(s'|s, a)}$: The state transition distribution. For each state $s \in S$ and action $a \in A$, it gives the probability that the system transitions to state $s' \in S$.
\item $P(R|s, a)$: The scalar reward distribution. For each state $s \in S$ reached by action $a \in A$, it defines a probability distribution across scalar rewards.
\end{itemize}

An episodic realization of an MDP proceeds as follows. First, an initial state $s_0 \in S$ is sampled from $D$. Then, the decision-making agent takes an action $a_0 \in A$. A new state $s_1 \in S$ is sampled from $P(s'|s_0, a_0)$. A reward $R_1$ is sampled from $P(R|s_0, a_0)$. This process repeats until a terminal state $s_N$ is reached. The reward of the full episode is $R=R_1 + R_2 + \cdots + R_N$.

The goal of the agent is to choose actions $a_0, a_1, \cdots$ to maximize the expected reward, $\mathbb{E}[R]$. There is a diverse set of problem setups, based on what information - if any - is available to the agent and the structure of the state and action spaces. For example, some approaches assume the agent has a probabilistic model of the state-transition distribution (model-based), and some do not (model-free). Since we do not have an explicit model of the system, our approach is model-free. Successful approaches to the aforementioned problem either try to compute the \textit{value} of each state (value-function methods) or directly try to optimize a decision-making strategy (policy-search methods). Because of the Markov property of MDPs, it suffices to choose each action as a function of the current state $s_t$ \cite{sutton1998reinforcement}. In this work, we take this approach and work directly in the space of policies.

A stochastic policy $\pi (a|s; \theta)$ defines a probability distribution over actions given states, where $\theta$ are the parameters of the distribution. We seek to find parameters $\theta$ which achieve the highest expected reward under a given MDP scenario, which can be framed as the following optimization problem:

\begin{equation}
\max_{\theta} \mathbb{E}[R|\pi_{\theta}]
\end{equation}

This optimization problem is explicitly solvable in the case where one knows the dynamics model and reward function, for example in the linear-quadratic case using the iLQR algorithm \cite{bertsekas1995dynamic}. However, in this work, we are considering the model-free case. If no model is known, the agent must act in the environment and use its experience to update its parameters and repeat this process over and over. This form of learning is known as policy gradient method, in which over each \textit{experience} the agent computes an approximation of the gradient $\nabla_{\theta} E[R|\pi_{\theta}]$ and updates its parameters.

To compute this gradient estimate, an estimator called the \textit{Score Function Gradient Estimator} (SFGE) is used. It is derived as follows:

\begin{equation} \label{eq:sfge}
\begin{aligned}
\nabla_{\theta} \mathbb{E}_x[f(x)] & = \nabla_{\theta} \int dx \,\, p(x|\theta) f(x) \\
& = \int dx \,\, \nabla_{\theta} p(x|\theta) f(x) \\
& = \int dx \,\, p(x|\theta) \frac{\nabla_{\theta} p(x | \theta)}{p(x|\theta)} f(x) \\
& = \int dx \,\, p(x|\theta) \nabla_{\theta} \log p(x|\theta) f(x) \\
& = \mathbb{E}_x[f(x)\nabla_{\theta}\log p(x|\theta)]
\end{aligned}
\end{equation}

The last equation gives us the unbiased estimator: Just sample $x_i \sim p(x|\theta)$ and compute $\hat{g_i} = f(x_i) \nabla_{\theta} \log p(x_i|\theta)$. Now if we assume $f$ to be our reward function and $s$ to be the whole trajectory $\tau = ((s_0, a_0, r_0), (s_1, a_1, r_2), \cdots)$, our gradient estimator becomes:

\begin{equation}
\begin{aligned}
\nabla_{\theta} \mathbb{E}_{\tau}[R] & = \mathbb{E}_{\tau}\left[R\nabla_{\theta} \sum \log \pi(a_t|s_t; \theta)\right] \\
& =\mathbb{E}\left[\sum_{t=0}^{T-1} \nabla_{\theta} \log \pi(a_t | s_t; \theta) \sum_{t'=t}^{T-1} r_{t'}\right]
\end{aligned}
\end{equation}

So a \enquote{vanilla} policy gradient algorithm follows:

\begin{algorithm}
\SetAlgoLined
\KwResult{Policy parameters $\theta$}
 Initialize policy parameters $\theta$\;
 \For{$i=1,2, \cdots$}{
  Collect a set of trajectories by executing the current policy\;
  At each timestep in the trajectory compute $\sum_{t'=t}^{T-1} r_{t'}$\;
  Update the policy using the average of the policy gradient estimates $\hat{g_i}$ for each trajectory\;
 }
 \caption{The vanilla policy gradient algorithm.}
\end{algorithm}

Since this section was brief, we point the reader to the following resources for more background. For a more in-depth look into MDPs, see~\cite{ng2003autonomous}. For a more in-depth look into reinforcement learning algorithms, see~\cite{sutton1998reinforcement}. For a more in-depth look into policy gradients, see~\cite{Schulman:EECS-2016-217}.

Reinforcement learning is a diverse field with a rich history. In the past, limited computation power and simple tabular algorithms restricted reinforcement learning to simple, low-dimensional tasks. Now, it has been combined with value functions, state models and policies approximated by high capacity models (e.g. neural networks), now known as deep reinforcement learning. Deep reinforcement learning has led to many recent successes, e.g. in helicopter flight~\cite{ng2003autonomous}, beating human experts in the game Go~\cite{silver2016mastering} and continuous control~\cite{lillicrap2015continuous}.

\section{Preliminary Analysis}\label{sec:preliminary analysis}

In this section, we explore how general machine learning algorithms perform on the standard wireless communication problem of modulation and demodulation. The question we seek to address is how well a generic gradient descent approach without any domain specific knowledge works on these tasks for which there exist well known solutions. By training a single learning agent together with a fixed conventional counterpart, we hope to get an estimate of the \textit{learnability} of modulation and demodulation before exploring the more generalized and complex collaborative setting.
% complex: no pun intended

\subsection{Single Agent Receiver} \label{sec:clustering}

The first question we would like to answer is how well an agent can recognize and demodulate an unknown signal based on a known bit sequence (\textit{preamble}) which is transmitted before the actual payload data. In contrast to prior work, which oftentimes considers parameter tuning across known modulation schemes or runs classification in a supervised learning setting, we assume no prior knowledge about common modulation schemes (QPSK, 8-PSK, 16-QAM, ...).

\subsubsection{Problem Statement}
The problem is formulated as follows: Given a transmitter with an arbitrary but fixed modulation scheme with up to 20 constellation points which transmits a known preamble and a large unknown batch of payload data across an AWGN channel - which bit-error rate can we achieve across the transmission of the payload under various noise conditions? Since we do not assume any knowledge about usual modulation schemes, the problems boils down to an unsupervised clustering problem. The algorithm has to find the number of clusters in the received samples, assign each cluster a mapping from samples to bit strings given the known preamble and then demodulate the payload with the learned demodulator. 

\subsubsection{Approach}
We use Lloyd's algorithm~\cite{kmeans} to solve a variant of k-means clustering called the \textit{jump method}, which is based on an information-theoretic approach to clustering presented in~\cite{distortion}. The algorithm runs k-means for every number of $k=1...N$ on the received sequence and every complex sample is interpreted as point in a 2D space. In our analysis, we limit $N$ to $N=20$. After running k-means for a fixed number of iterations, the algorithm calculates the \textit{minimum average distortion} $d_k$ of the clustering.

\begin{definition}{Minimum average distortion:}
Let X be a two-dimensional random variable representing the real and imaginary part of the received samples. Let the means $c_1 \cdots c_N$ be given and let $c_x$ be the closest mean to a given sample of X under L2 norm. Then the minimum average distortion $d[k]$ is defined as minimum variance across all possible clusterings:

\begin{equation}
d[k] = \frac{1}{2} min_{c_1...c_k} E[(X-c_x)^T(X-c_x)]
\end{equation}

In the setting with a finite training sequence, the expectation gets replaced by the sample mean across all training samples.
\end{definition}

Once all values of the distortion function have been collected, the algorithm computes the jump function $J(k)_{k=1...N}$, which is the discrete derivative of the inverse of the distortion function:

\begin{equation}
J(k) = d(k)^{-1} - d(k-1)^{-1}
\end{equation}

The algorithm then uses the value of $k$ which maximizes $J(k)$ as $k$ in k-means clustering to return $k$ cluster means. Once we have found a clustering, we can label each cluster with its most common bit string across all the points within the cluster, thus creating a mapping from complex values back to bit strings. 

The logic behind this algorithm becomes apparent if we think about how the distortion develops if we increase $k$. Assuming the k-means algorithm converges, the distortion will always decrease with higher k because more clusters can explain the variance in the dataset better than fewer ones. However, once we increase k beyond the actual number, splitting the clusters will only marginally decrease the distortion. Thus, we are looking for the \enquote{jump} in the transformed distortion function $d(k)^{-1}$, which points us to the number of clusters beyond which diving the dataset further does not explain much more variance. More information and mathematical support for this algorithm in the form of asymptotic reasoning can be found in~\cite{distortion}.

In general, the initialization for the k-means algorithm plays an important role. We use a variant of the k-means++ scheme~\cite{kmeans++} but instead of sampling initial means in a probabilistic manner we simply choose the points that maximize the distance to any of the previously chosen initial means.

\subsubsection{Evaluation}
In our analysis, we examine how well our clustering approach works on data modulated with common modulation schemes. We first send a preamble to create a clustering and a mapping of cluster means back to bits. Then, a large chunk of data is transmitted to evaluate the bit-error rate with the learned demodulator. Besides the BER, we also record the number of clusters which the algorithm identified and plot both measures over different noise intensities. Table~\ref{tab:params} shows the hyperparameters used during training.

\begin{table}[!ht]
\vspace{.5cm}
\fontsize{9}{9}\selectfont
\renewcommand{\arraystretch}{1.3}
\centering
\begin{tabular}{c c}
\hline
\textbf{Hyperparameter} & \textbf{Value} \\
\hline\hline
Number of iterations for k-means algorithm & 50 \\
Number of symbols for testing & $10^7$ \\
Number of constellation points & 16 \\
\end{tabular}
\caption{Hyperparameters of the unsupervised clustering algorithm test.}
\label{tab:params}
\end{table}

We focus our evaluation on the case where the transmitter sends data modulated with 16-QAM. Among the modulation schemes of order four or smaller, 16-QAM is the most complex and hence interesting to analyze and we found that the results for BPSK, QPSK and 8-PSK deliver similar results. Using a standard modulation scheme allows us to compare the performance of our algorithm to that of a conventional decoder as baseline, which assumes full knowledge about the modulation and thus serves as a statistical upper bound.

\begin{figure}[H]
\centering
% This file was created by matplotlib2tikz v0.6.6.
\begin{tikzpicture}

\definecolor{red}{rgb}{1,0.254901960784314,0.211764705882353}
\definecolor{blue}{rgb}{0,0.454901960784314,0.850980392156863}
\definecolor{yellow}{rgb}{1,0.862745098039216,0}
\definecolor{green}{rgb}{0.23921568627451,0.6,0.43921568627451}
\definecolor{purple}{rgb}{0.52156862745098,0.0784313725490196,0.294117647058824}

\begin{axis}[
xlabel={$E_b/N_0$ [dB]},
ylabel={Bit-error rate (BER)},
xmin=0, xmax=16,
ymin=1e-07, ymax=1,
ymode=log,
axis line on top,
width=3.1in,
height=2in,
grid=both,
xtick={0,2,4,6,8,10,12,14,16},
ytick={1e-08,1e-07,1e-06,1e-05,0.0001,0.001,0.01,0.1,1,10},
yticklabels={0.0,$ 10^{-7}$,$ 10^{-6}$,$ 10^{-5}$,$ 10^{-4}$,$ 10^{-3}$,$ 10^{-2}$,$ 10^{-1}$,$ 10^{0}$,},
xtick pos=both,
ytick pos=left,
legend entries={{Baseline},{Clustering},{\#Clusters}},
legend style={at={(0.5,1)},
anchor=south,
draw=none,
%text height=0.7ex
},
legend columns=3,
legend cell align={left}
]
\path [draw=red, fill=red, opacity=0.35] (axis cs:0,0.22056556525)
--(axis cs:0,0.16674312975)
--(axis cs:0.410256,0.1727238496)
--(axis cs:0.820513,0.1531841554)
--(axis cs:1.230769,0.18802276465)
--(axis cs:1.641026,0.15857213425)
--(axis cs:2.051282,0.13188105985)
--(axis cs:2.461538,0.13427542385)
--(axis cs:2.871795,0.1011380666)
--(axis cs:3.282051,0.1010707333)
--(axis cs:3.692308,0.06992789655)
--(axis cs:4.102564,0.05822095565)
--(axis cs:4.512821,0.05849786835)
--(axis cs:4.923077,0.0514146123)
--(axis cs:5.333333,0.0521312843)
--(axis cs:5.74359,0.03960929275)
--(axis cs:6.153846,0.0292539163)
--(axis cs:6.564103,0.0250773599)
--(axis cs:6.974359,0.0206028234)
--(axis cs:7.384615,0.0148870033)
--(axis cs:7.794872,0.01098267025)
--(axis cs:8.205128,0.0088984189)
--(axis cs:8.615385,0.0064114692)
--(axis cs:9.025641,0.00404924245)
--(axis cs:9.435897,0.0031946019)
--(axis cs:9.846154,0.00219430195)
--(axis cs:10.25641,0.00149234285)
--(axis cs:10.666667,0.00088982915)
--(axis cs:11.076923,0.0005383)
--(axis cs:11.487179,0.000307477810443694)
--(axis cs:11.897436,0.000167882571709099)
--(axis cs:12.307692,8.71011277882597e-05)
--(axis cs:12.717949,4.26902877509845e-05)
--(axis cs:13.128205,1.97791306373515e-05)
--(axis cs:13.538462,8.14505682552077e-06)
--(axis cs:13.948718,3.34727062291366e-06)
--(axis cs:14.358974,1.35958456608574e-06)
--(axis cs:14.769231,4.4788490752043e-07)
--(axis cs:15.179487,1.10370996007421e-07)
--(axis cs:15.589744,2.62599262652488e-08)
--(axis cs:16,2.704627669473e-10)
--(axis cs:16,1.02704627669473e-08)
--(axis cs:16,1.02704627669473e-08)
--(axis cs:15.589744,4.37400737347513e-08)
--(axis cs:15.179487,1.44629003992579e-07)
--(axis cs:14.769231,4.9711509247957e-07)
--(axis cs:14.358974,1.50041543391426e-06)
--(axis cs:13.948718,3.59272937708634e-06)
--(axis cs:13.538462,8.58494317447923e-06)
--(axis cs:13.128205,2.05308693626485e-05)
--(axis cs:12.717949,4.42647122490155e-05)
--(axis cs:12.307692,9.18388722117403e-05)
--(axis cs:11.897436,0.000176317428290901)
--(axis cs:11.487179,0.000321192189556306)
--(axis cs:11.076923,0.000586525)
--(axis cs:10.666667,0.00109351585)
--(axis cs:10.25641,0.00190528215)
--(axis cs:9.846154,0.00249146805)
--(axis cs:9.435897,0.0034438031)
--(axis cs:9.025641,0.00889384255)
--(axis cs:8.615385,0.0068668608)
--(axis cs:8.205128,0.0099890061)
--(axis cs:7.794872,0.01613016475)
--(axis cs:7.384615,0.0163328717)
--(axis cs:6.974359,0.0267839216)
--(axis cs:6.564103,0.0305562351)
--(axis cs:6.153846,0.0374178387)
--(axis cs:5.74359,0.04888683225)
--(axis cs:5.333333,0.0689319657)
--(axis cs:4.923077,0.0569613977)
--(axis cs:4.512821,0.06686984165)
--(axis cs:4.102564,0.12154261935)
--(axis cs:3.692308,0.12974789845)
--(axis cs:3.282051,0.1766421367)
--(axis cs:2.871795,0.1310053334)
--(axis cs:2.461538,0.21014374615)
--(axis cs:2.051282,0.19781713515)
--(axis cs:1.641026,0.23568513575)
--(axis cs:1.230769,0.26351185035)
--(axis cs:0.820513,0.2098613046)
--(axis cs:0.410256,0.2349475304)
--(axis cs:0,0.22056556525)
--cycle;

\addplot [very thick, blue]
table {%
0 0.140899475
0.410256 0.13178885
0.820513 0.122854125
1.230769 0.113969525
1.641026 0.1052297
2.051282 0.0966184
2.461538 0.088169925
2.871795 0.079948475
3.282051 0.071942925
3.692308 0.064208525
4.102564 0.056784975
4.512821 0.049729025
4.923077 0.04308635
5.333333 0.0368782
5.74359 0.0311724
6.153846 0.02596535
6.564103 0.02130545
6.974359 0.017179875
7.384615 0.013608175
7.794872 0.01057245
8.205128 0.00801805
8.615385 0.00592955
9.025641 0.004276975
9.435897 0.0029977
9.846154 0.0020424
10.25641 0.001341525
10.666667 0.00084555
11.076923 0.0005079
11.487179 0.00029255
11.897436 0.0001594
12.307692 8.2425e-05
12.717949 4.0325e-05
13.128205 1.8575e-05
13.538462 7.475e-06
13.948718 3.1e-06
14.358974 1.2e-06
14.769231 4e-07
15.179487 1.25e-07
15.589744 5e-08
16 0
};
\addplot [very thick, red]
table {%
0 0.1936543475
0.410256 0.20383569
0.820513 0.18152273
1.230769 0.2257673075
1.641026 0.197128635
2.051282 0.1648490975
2.461538 0.172209585
2.871795 0.1160717
3.282051 0.138856435
3.692308 0.0998378975
4.102564 0.0898817875
4.512821 0.062683855
4.923077 0.054188005
5.333333 0.060531625
5.74359 0.0442480625
6.153846 0.0333358775
6.564103 0.0278167975
6.974359 0.0236933725
7.384615 0.0156099375
7.794872 0.0135564175
8.205128 0.0094437125
8.615385 0.006639165
9.025641 0.0064715425
9.435897 0.0033192025
9.846154 0.002342885
10.25641 0.0016988125
10.666667 0.0009916725
11.076923 0.0005624125
11.487179 0.000314335
11.897436 0.0001721
12.307692 8.947e-05
12.717949 4.34775e-05
13.128205 2.0155e-05
13.538462 8.365e-06
13.948718 3.47e-06
14.358974 1.43e-06
14.769231 4.725e-07
15.179487 1.275e-07
15.589744 3.5e-08
16 5e-09
};

%dummy plot
\addplot [very thick, yellow]
table {%
-20 0
-20 1
};
\end{axis}

\begin{axis}[
ylabel={Number of Clusters},
xmin=0, xmax=16,
ymin=2, ymax=24,
axis line on top,
width=3.1in,
height=2in,
%axis y line=right,
ytick={0,4,8,12,16,20},
xmajorticks=false,
ytick pos=right
]
\path [draw=yellow, fill=yellow, opacity=0.35] (axis cs:0,17.06694755225)
--(axis cs:0,11.13305244775)
--(axis cs:0.410256,8.47127999195)
--(axis cs:0.820513,10.61084402945)
--(axis cs:1.230769,5.461279631)
--(axis cs:1.641026,7.31801948465)
--(axis cs:2.051282,9.1325582443)
--(axis cs:2.461538,8.07697955295)
--(axis cs:2.871795,12.60185280545)
--(axis cs:3.282051,10.0476130853)
--(axis cs:3.692308,13.7026585413)
--(axis cs:4.102564,12.80277970945)
--(axis cs:4.512821,15.9250514423)
--(axis cs:4.923077,16.33333333335)
--(axis cs:5.333333,13.95724320785)
--(axis cs:5.74359,15.71180582705)
--(axis cs:6.153846,15.9909317538)
--(axis cs:6.564103,16.1754311627)
--(axis cs:6.974359,15.95993827515)
--(axis cs:7.384615,16.6622024821)
--(axis cs:7.794872,16.4622024821)
--(axis cs:8.205128,16.883772234)
--(axis cs:8.615385,16.602785537)
--(axis cs:9.025641,16.11695410845)
--(axis cs:9.435897,16.40559468115)
--(axis cs:9.846154,16.40559468115)
--(axis cs:10.25641,16.50139050015)
--(axis cs:10.666667,16.2928932188)
--(axis cs:11.076923,16.1464466094)
--(axis cs:11.487179,15.9891814893)
--(axis cs:11.897436,15.941886117)
--(axis cs:12.307692,15.941886117)
--(axis cs:12.717949,16)
--(axis cs:13.128205,16)
--(axis cs:13.538462,16)
--(axis cs:13.948718,16)
--(axis cs:14.358974,16)
--(axis cs:14.769231,16)
--(axis cs:15.179487,16)
--(axis cs:15.589744,16)
--(axis cs:16,16)
--(axis cs:16,16)
--(axis cs:16,16)
--(axis cs:15.589744,16)
--(axis cs:15.179487,16)
--(axis cs:14.769231,16)
--(axis cs:14.358974,16)
--(axis cs:13.948718,16)
--(axis cs:13.538462,16)
--(axis cs:13.128205,16)
--(axis cs:12.717949,16)
--(axis cs:12.307692,16.258113883)
--(axis cs:11.897436,16.258113883)
--(axis cs:11.487179,16.4108185107)
--(axis cs:11.076923,16.8535533906)
--(axis cs:10.666667,17.7071067812)
--(axis cs:10.25641,17.69860949985)
--(axis cs:9.846154,17.19440531885)
--(axis cs:9.435897,17.19440531885)
--(axis cs:9.025641,17.08304589155)
--(axis cs:8.615385,17.597214463)
--(axis cs:8.205128,17.516227766)
--(axis cs:7.794872,17.3377975179)
--(axis cs:7.384615,17.5377975179)
--(axis cs:6.974359,17.04006172485)
--(axis cs:6.564103,17.6245688373)
--(axis cs:6.153846,17.4090682462)
--(axis cs:5.74359,17.48819417295)
--(axis cs:5.333333,16.44275679215)
--(axis cs:4.923077,17.66666666665)
--(axis cs:4.512821,17.2749485577)
--(axis cs:4.102564,16.79722029055)
--(axis cs:3.692308,18.2973414587)
--(axis cs:3.282051,15.7523869147)
--(axis cs:2.871795,16.19814719455)
--(axis cs:2.461538,13.72302044705)
--(axis cs:2.051282,14.8674417557)
--(axis cs:1.641026,13.68198051535)
--(axis cs:1.230769,11.138720369)
--(axis cs:0.820513,16.38915597055)
--(axis cs:0.410256,14.72872000805)
--(axis cs:0,17.06694755225)
--cycle;

\addplot [very thick, yellow]
table {%
0 14.1
0.410256 11.6
0.820513 13.5
1.230769 8.3
1.641026 10.5
2.051282 12
2.461538 10.9
2.871795 14.4
3.282051 12.9
3.692308 16
4.102564 14.8
4.512821 16.6
4.923077 17
5.333333 15.2
5.74359 16.6
6.153846 16.7
6.564103 16.9
6.974359 16.5
7.384615 17.1
7.794872 16.9
8.205128 17.2
8.615385 17.1
9.025641 16.6
9.435897 16.8
9.846154 16.8
10.25641 17.1
10.666667 17
11.076923 16.5
11.487179 16.2
11.897436 16.1
12.307692 16.1
12.717949 16
13.128205 16
13.538462 16
13.948718 16
14.358974 16
14.769231 16
15.179487 16
15.589744 16
16 16
};
\end{axis}

\end{tikzpicture}
  \caption{Performance comparison of demodulation with unsupervised clustering (red) assuming a known preamble of 1000 symbols versus a conventional decoder (baseline) for 16-QAM. The yellow line shows the algorithm's estimate of the number of constellation points in the data.}
  \label{fig:ber_num_qam}
%\vspace{-1.5cm}
\end{figure}
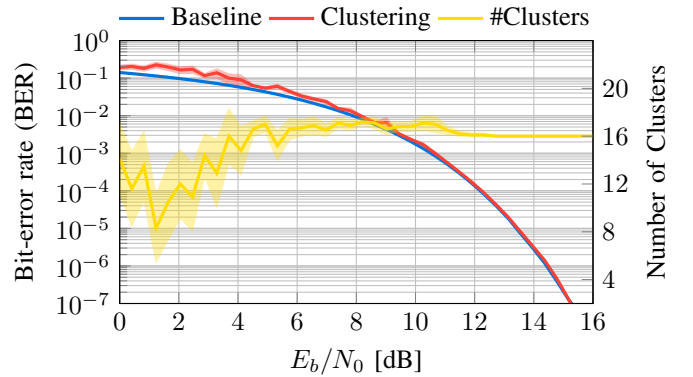

\begin{figure}[!ht]
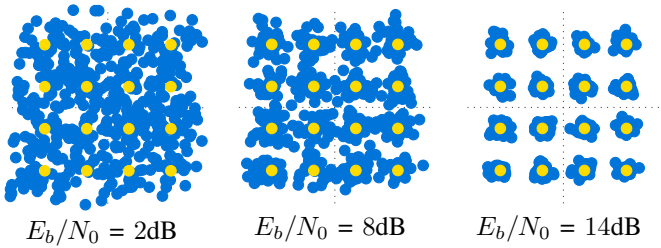

\begin{subfigure}{0.165\textwidth}
\centering
\input{2db.tex}
$E_b/N_0$ = \si{2}{dB}
\label{fig:2db}
\end{subfigure}%
\begin{subfigure}{0.165\textwidth}
\centering
\input{8db.tex}
$E_b/N_0$ = \si{8}{dB}
\label{fig:8db}
\end{subfigure}%
\begin{subfigure}{0.165\textwidth}
\centering
\input{tikz/14db.tex}
$E_b/N_0$ = \si{14}{dB}
\label{fig:14db}
\end{subfigure}%
\caption{Visualization of the noise power. The yellow points are the originally sent constellation points. The blue points represent the input to the clustering algorithm.}
\label{fig:noises}
\end{figure}

Figure~\ref{fig:ber_num_qam} shows the performance in terms of BER and number of identified clusters for a transmission of 16-QAM-modulated data. For high $E_b/N_0$ ratios (i.e. low noise) of $>$\si{10}{dB}, the learned decoder matches the BER performance of the baseline. Although the algorithm performs very poorly on identifying the correct number of clusters in high noise situations, the resulting BER is still very close to the ideal solution and does not break down. Since the bit mappings of the constellation points are spatially correlated, our algorithm gets approximately the same number of bits right even when mapping multiple clusters to a single point. Thus, it does not matter that the algorithm underestimates the number of clusters and confuses constellation points, because the resulting error is close to its statistical expectation. Figure~\ref{fig:noises} gives some intuition for different noise situations in terms of the $E_b/N_0$ ratio.

\begin{figure}[!ht]
\centering
\input{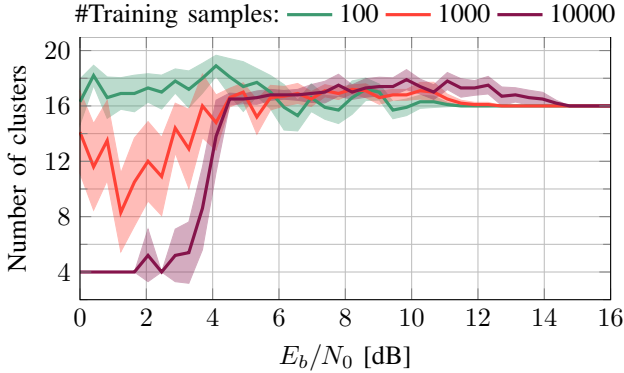}
  \caption{Number of identified clusters over different noise levels, parametrized by the length of the training preamble.}
  \label{fig:num_qam}
\end{figure}

Figure~\ref{fig:num_qam} shows the development of the number of identified clusters over time for different preamble lengths. A high number of training samples leads the algorithm to underestimate the number of clusters because the constellations points in the middle of 16-QAM are indistinguishable while the noisy quadratic shape of the constellation diagram is best explained by four clusters. When training with fewer samples the algorithm slightly overestimates the number of clusters because the number of samples per cluster is low. However, in comparison to the other runs it converges the fastest. 

\begin{figure}[!ht]
\centering
\input{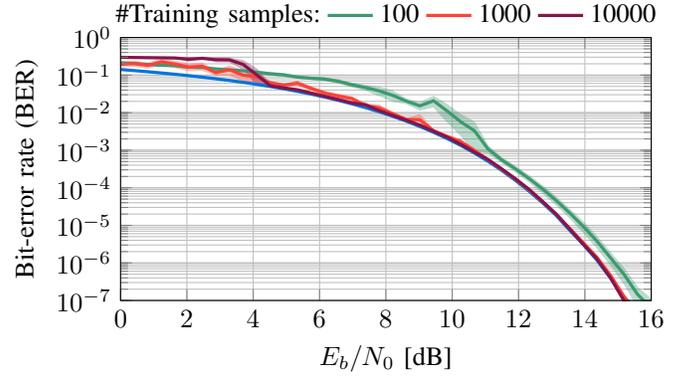}
  \caption{Bit-error rate over different noise levels, parametrized by the length of the preamble used during training. The blue curve shows the performance of a standard coherent 16-QAM demodulator (baseline).}
  \label{fig:ber_qam}
  \vspace{-.5cm}
\end{figure}

Figure~\ref{fig:ber_qam} shows the bit-error rate for schemes trained with different preamble lengths. The algorithm trained on a long preamble does very poorly in presence of high noise because of the underestimation of the number of clusters. However, once the $E_b/N_0$ ratio increases past \si{4}{dB}, the algorithm quickly converges to the performance of the baseline. The algorithm trained on very few examples, despite giving the best estimate of the number of clusters, never reaches baseline performance because the small number of samples leads to an imprecise guess of the cluster means.

As a result, we can conclude that simple non-parametric and unsupervised clustering does very well on demodulating received signals based on a known preamble in comparison to standard coherent demodulators\footnote{While we also tried using spectral clustering~\cite{clustering} based on various adjacency graphs, we found that this method was generally less capable.}. When we later move to the multi-agent setting, there is obviously no need for neural architectures or sophisticated learning algorithms for the sole purpose of demodulation when there is a shared preamble.

\subsection{Single Agent Transmitter}\label{subsec:single}
We now turn around the setting and ask how well an agent can learn the functionality of a modulator and communicate with a given receiver. The receiver is static and will try to demodulate the signal based on a common modulation scheme, however we assume no knowledge about common modulation schemes in the transmitter. The goal of the transmitter is to minimize the expected bit error rate of the receiver.

\subsubsection{Problem Statement} The transmitter is given a random but fixed binary preamble $b$ and consecutively maps fixed-length bit strings $b_i$ of the preamble to complex symbols $o(b_i)$. The symbols get send over an AWGN channel to the receiver, who receives $\hat{o}(b_i) = o(b_i) + n$ with $n\sim\mathcal{CN}(0, N_0)$. The receiver then demodulates the noisy signals using its fixed modulation scheme and recovers $\hat{b}$. Finally, the transmitter receives a reward based on the difference between the preamble it sent and the bit sequence that the receiver decoded.

\begin{figure}[!ht]
\centering
\includegraphics[width=3.5in]{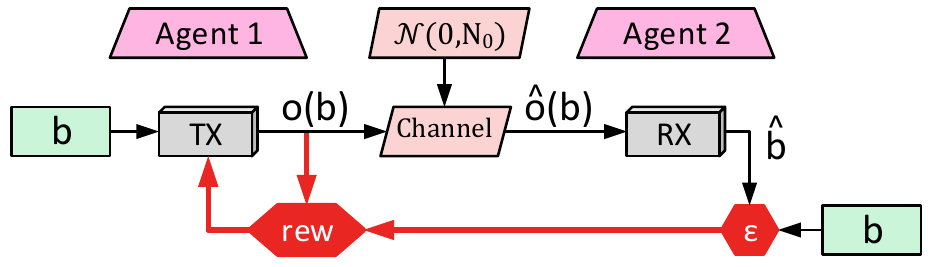}
  \caption{The learning transmitter tries to communicate with a static receiver. The learning process is centralized because the reward is calculated based on the output of the receiver.}
  \label{fig:single_transmitter}
\end{figure}

\subsubsection{Approach} 
\begin{figure}[!ht]
\centering
\includegraphics[width=2.4in]{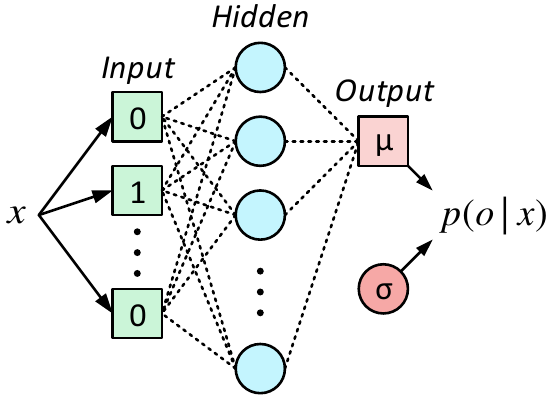}
  \caption{Architecture of the transmitter. A fixed number of bits is used as input into a fully connected neural network which outputs a real and imaginary mean $\mu = [Re\{\mu\}, Im\{\mu\}]$. The complex output symbols are then sampled i.i.d. from a multivariate Gaussian distribution parametrized by the real and imaginary means and a separate variable $\sigma = [Re\{\sigma\}, Im\{\sigma\}]$ representing the standard deviations of the distribution.}
  \label{fig:transmitter}
\end{figure}

The transmitter is parametrized by a neural network that takes as input a single bit string of given length and outputs a single complex symbol. It uses the bipolar representation of $-1$ and $1$ to represent bits. The network has one fully connected hidden layer that uses the ReLU activation function and its hidden units are fully connected to a single neuron which calculates  $\mu = [Re\{\mu\}, Im\{\mu\}]$. To encourage exploration, the output of the transmitter is drawn from a complex normal distribution parameterized by the output $\mu$ and a trainable standard deviation $\sigma = [Re\{\sigma\}, Im\{\sigma\}]$, which can be seen as just another variable weight of the neural network. Figure~\ref{fig:transmitter} visualizes the architecture of the transmitter.

The agent receives a loss signal from the environment which is equal to the L1-norm between the sent preamble $b$ and the preamble $\hat{b}$ recovered by the receiver, plus a factor corresponding to the energy of the complex symbols outputted by the transmitter:

\begin{equation}\label{eq:loss}
    L_i = \lVert b_i - \hat{b_i}\rVert_1 + \lambda_p\lVert o(b_i) \rVert_2^2.
\end{equation}

This allows the network to optimize for minimizing the bit error rate while also trading off for the energy of the signal. The negative loss is then used as the advantage estimator for computing the SFGE of the expected reward. This gradient is used in conjunction with the Adam optimizer \cite{kingma2014adam} to train the network using the vanilla policy gradient algorithm described in Section~\ref{subsec:rlpolicy}. In contrast to how the policy gradient algorithm was described before, here our episode length is $1$. Each input, transmission, output and associated reward is taken to be a single episode of training. This is necessary because there is no temporal structure to the actions taken. The state in this case is the past history of transmissions and rewards but it is not modeled explicitly by the network. The gradient estimator then, is simply

\begin{equation}
    \nabla_{\theta} \mathbb{E}[R] \approx -\sum_{i=1}^N L_i \nabla_{\theta} \log p(o(x_i)|x_i)
\end{equation}

\subsubsection{Results} 
\begin{figure}[!ht]
\centering
\input{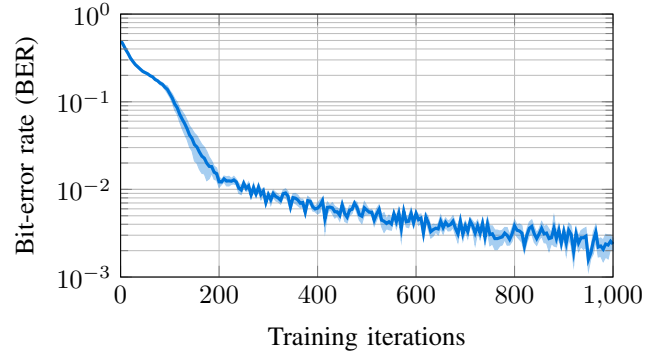}
  \caption{Bit-error rate over number of training iterations of the system with a learning transmitter and a fixed 16-QAM receiver. Hyperparameters: Initial $\log(\sigma)=-2$. For the other parameters refer to Table~\ref{tab:params_1}.}
  \label{fig:loss_single}
\end{figure}

We found that the network was capable of learning approximations of QPSK, 8-PSK and 16-QAM for two, three and four bit transmissions respectively with reasonable preambles lengths of a few hundred symbols within a few hundred iterations. Figure~\ref{fig:loss_single} shows a typical learning curve for this set-up given a fixed 16-QAM receiver. The learning progress is quantified by the bit-error rate of the preamble transmission and shows a rapid decrease of the BER within the first 200 iterations, because the network explores positions for the constellation points which minimize the error rate. After that, the transmitter has learned to put the constellation points in the right position as dictated by the receiver. Subsequently, it starts to exploit the learned constellation by reducing the standard deviation of the sampling process, which leads to a slow but steady decrease in BER. These results confirm that the loss signal is expressive enough for the network to learn an intelligible communication scheme. 

\section{Main Problem Formulation}\label{sec:main-problem-formulation}

After studying how agents can learn to receive and transmit given a static counterpart, it an obvious question to ask whether it is possible to combine both aspects and train a receiver and a transmitter simultaneously. As previously mentioned, we intend to solve the problem in which two agents are only allowed to communicate strictly through an unknown noisy channel using their respective learned digital communication scheme. Unlike previous work in this area, which significantly relaxed the problem by allowing for shared weights, shared gradients or a fully differentiable communication process, it is no longer clear how the reward signal should be computed and how it will be communicated between agents.

\section{Main Problem Setup}\label{sec:main-problem-setup}

We propose the following approach as a solution for the problem described above. Two actors, Agent 1 and Agent 2, share a fixed bit string $b$ that acts as the preamble. Each agent contains both a transmitter and a receiver. The transmitters are an instance of the neural network from Section~\ref{subsec:single} that takes a bit string as input and outputs a complex number which represents the actor's transmission for that bit string. The receiver runs k-nearest neighbors (kNN) on each complex number by comparing it with the rest of the modulated preamble and generates a guess for each transmission. The noise function of the channel is parametrized by $n\sim\mathcal{CN}(0,N_0)$.

\begin{figure}[!ht]
\centering
\includegraphics[width=3.5in]{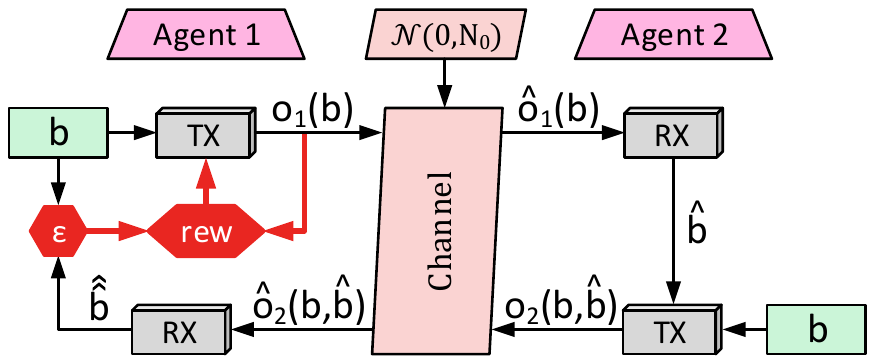}
  \caption{Full system with decentralized reward. Agent 1 and 2 share a fixed preamble $b$, echo it back and forth over the noisy channel and use the difference between the sent and received preamble to compute the reward \textit{rew}.}
  \label{fig:system}
    \vspace{-.2cm}
\end{figure}

During each iteration, each actor takes turns in completing an entire update operation for their transmitter. The update operation is described as follows:
\\
\begin{enumerate}
    \item Agent 1 modulates the preamble to produce the signal $o_1(b)$ and sends it over the channel.
    \item Agent 2 receives the noisy modulated signal $\hat{o}_1(b)=o_1(b)+n$ and produces a guess of the preamble $\hat{b}$ using its receiver.
    \item Agent 2 modulates both the preamble $b$ and its guess of the received preamble $\hat{b}$ to produce the signal $o_2(b,\hat{b})$ and sends it over the channel.
    \item Agent 1 receives the noisy modulated signal $\hat{o}_2(b,\hat{b})$ and computes a guess of $\hat{b}$, which is $\hat{\vphantom{\rule{1pt}{8.5pt}}\smash{\hat{b}}}$, with its receiver using the modulation of $b$ as reference.
    \item Agent 1 then computes a loss function given $b$ and $\hat{\vphantom{\rule{1pt}{8.5pt}}\smash{\hat{b}}}$ and updates its parameters using policy gradients.
    \item Agent 1 and Agent 2 switch roles and the loop starts over at step 1.
\end{enumerate}

The receiver has two distinct behaviors depending on the type of signal it receives. If it receives only the modulated preamble, then it will demodulate each symbol with the help of the other known modulated symbols of the preamble. If instead it receives the modulated preamble along with another modulated signal, it will demodulate the latter signal by finding the nearest neighbors in the former. 

The secondary behavior of the receiver is to combat the noise that Agent 2's transmitter adds to $\hat{b}$. The modulation of the preamble gives Agent 1's receiver a noisy idea of the modulation scheme used by Agent 2's transmitter, so that the receiver can produce an educated demodulation of $\hat{o}_2(\hat{b})$. 

The loss function is defined to be a weighted combination of the bit error rate and average symbol energy as in Equation~\ref{eq:loss}. The loss is then used in computing the policy gradient as described in Algorithm~\ref{alg:full-learning}. 

\begin{algorithm}
\SetAlgoLined
\KwResult{Transmitter parameters $\theta_1$ and $\theta_2$}
 Initialize policy parameters $\theta_1$ and $\theta_2$\;
 \While{stopping condition not met}{
  Agent 1 sends b to Agent 2 and receives an echoed version $\hat{\vphantom{\rule{1pt}{8.5pt}}\smash{\hat{b}}}$\;
  Agent 1 calculates the reward for each symbol, the negative loss $R_i = -L_i$\;
  Agent 1 calculates $\nabla_{\theta_1} \mathbb{E}[R] \approx - \sum_{i=1}^N L_i \nabla_{\theta_1} \log p_1(o(x_i)|x_i)$\;
  Agent 1 performs a gradient update on its parameters\;
  Agent 1 and 2 switch roles\;
 }
 \caption{Decentralized learning of modulation schemes.}
 \label{alg:full-learning}
\end{algorithm}

The quality of the gradient updates for each agent is dependent on the quality of the other agent's transmitter. Because of this interdependency, the training is a complex optimization problem in which the signal is constantly changing due to both internal and external updates. Thus exploration plays an important role in the learning process and we investigate the visualization and interpretation of these dynamics in the next section. 

\section{Results}\label{sec:results}

In this section, we describe the results of a thorough analysis of our approach. First, we comment on the training behavior, which reveals some very interesting strategies to learn a robust modulation scheme. Next, we qualitatively describe the influence of each of the hyperparameters on the learning progress and on the final scheme. Finally, we give quantitative results on the performance of the learned modulation schemes. All quantitative results have been collected and are depicted in terms of their mean and standard deviation across ten runs with different random seeds.

\subsection{Training Behavior}

\begin{figure}[!ht]
\vspace{0.2cm}
\input{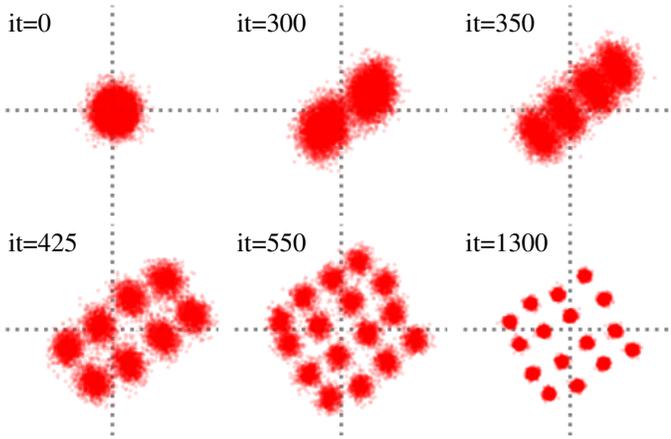}
\caption{Sampled output of the transmitter during training without symbol energy limit. See Appendix~\ref{hyps} Table~\ref{tab:params_1} for hyperparameters.}
\label{fig:training1}
\end{figure}

For brevity, we restrict our analysis the most complex case within the set of modulation schemes of order four or lower and thus focus on schemes with 16 constellations points. We provide both agents with a preamble of set length $M$ where $M$ is a hyperparameter. As described above, the agents take turns in transmitting and receiving preambles and echos through the AWGN channel and we record the loss of each agent's transmitter at every time step. Figure~\ref{fig:training1} shows the development of the modulation scheme of a given transmitter with the set of hyperparameters listed in Table~\ref{tab:params_1}. These plots were produced by having the transmitter modulate a sequence of bit strings with the modulation scheme it has learned up until a certain number of iterations and recording the complex outputs. Each red point represents the modulation of a single bit string, and the overall plot visualizes the means and standard deviations of the network at that iteration\footnote{Note that the randomness in the plots does not stem from the AWGN channel, since the plots are constructed from the outputs of the transmitter. Instead, the point clouds demonstrate the stochasticity of the neural network's outputs.}.

From the graphs it is clear that the training is broken up into distinct phases, ultimately converging to an optimal solution which represents a rotated version of 16-QAM. Initially, all of the constellation points are near the origin and have high variance. The random initialization of the network's weights and biases breaks any initial symmetry. After a few hundred iterations, the network splits the points into two clusters of eight constellation points each. Next, the network splits the points into four clusters of four constellation points each on a line that goes through the origin. After around 400 iterations, it begins to split the points into eight clusters of two constellation points, each in a direction perpendicular to the original cluster direction. After splitting each of the clusters one more time, each constellation point has its individual position in the constellation diagram. Finally, the network reduces the standard deviation on the transmitted points and converges on an optimal scheme which is very similar to a rotated version of 16-QAM. 

One other noteworthy observation is that, upon splitting, the agent reliably chooses every subset of constellations points in a way that minimizes the Hamming distance between the points in the subset, resulting in a steadily decreasing BER throughout training. In transmitting four bits, the network learns groups of four constellation points that are internally Gray coded but there is no global Gray coding across the whole constellation. This is likely due to the fact that locally Gray coded solutions represent local minima and a global Gray code is a non trivial combinatorial problem. Because we did not implement any techniques for escaping these minima such as simulated annealing, it is reasonable for the network to generally fall into locally Gray coded solutions.  

\begin{figure}[!ht]
\centering
\input{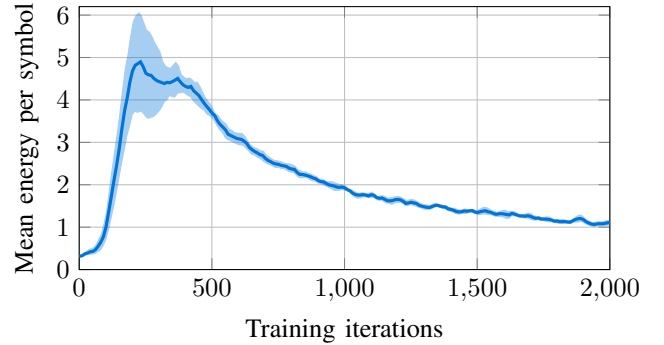}
  \caption{Development of the mean energy per symbol $\bar{E_s}$ over the course of the training with unrestricted energy.}
  \label{fig:unrestricted}
\end{figure}

If the energy per symbol is not restricted but only penalized with a soft loss, the network will initially increase the average energy beyond the final constellation to allow for greater exploration in the mapping. This gives the network enough room to split the clusters under any noise situation without increasing the bit error rate. After it achieves a reasonably low bit rate by splitting the clusters and moving the constellation points, it will begin to reduce the symbol energy and find a pareto-optimal point in the BER versus energy trade-off. Figure~\ref{fig:unrestricted} shows how the mean energy per symbol develops over training time.

\begin{figure}[!ht]
\input{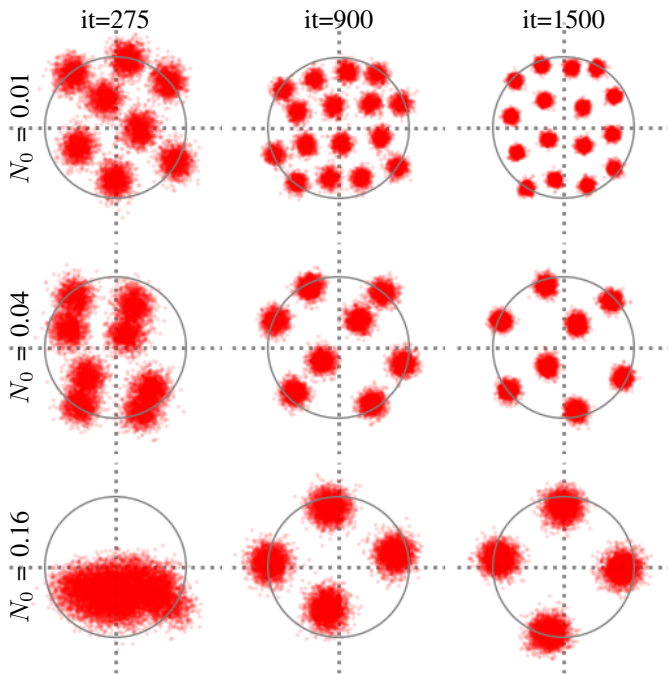}
\caption{Sampled output of the transmitter with restricted energy (unit circle) at different iterations (columns) and noise power densities $N_0$ (rows). See appendix~\ref{hyps} Table~\ref{tab:params_2} for the other hyperparameters.}
\label{fig:training3}
\end{figure}

In contrast, Figure~\ref{fig:training3} shows result for the case that the energy of the output signals is restricted. If the maximum mean symbol energy across all constellation points is hard-clamped to the unit circle, the network cannot simply increase the output energy to counter the noise anymore. Conventional radio system would usually decrease the order of the modulation scheme in the presence of higher noise e.g. from 16-QAM to QPSK. However, because we fixed the length of each input to four bits, the network has no other choice but to arrange all possible 16 constellation points in some way.

In Figure~\ref{fig:training3}, we see that increasing the noise power density steadily reduces the number of indistinguishable clusters from 16 individual points over eight clusters with two constellation points each down to only four clusters with four constellation points each when the noise is especially harsh. What is interesting about this is that the network chooses to \enquote{sacrifice} the same bit across all clusters, and to minimize hamming distance for the rest of the bits between the clusters. Thus, the network inherently learns the concept of decreasing the order of the modulation scheme in the presence of noise. This result is especially surprising because the system cannot exploit the lower order scheme to reduce the BER because it is forced to send four bits in each symbol. Section~\ref{sec:quant} will present some insight on why the networks behaves in the presented fashion nonetheless.

\begin{figure}[!ht]
\centering
\input{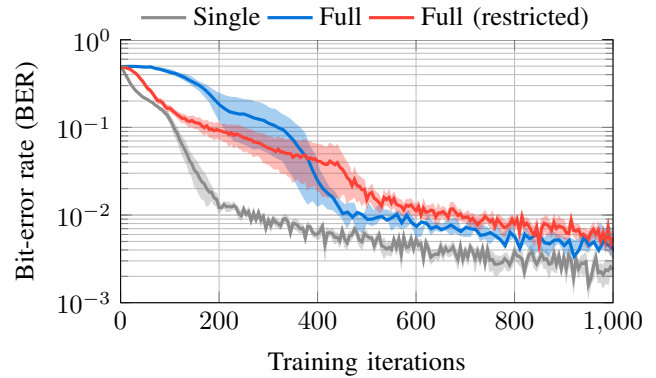}
  \caption{Bit-error rate over the first 1000 iterations. Hyperparameters: Initial $\log(\sigma)=-2$ for the single transmitter and the full system with restricted energy. For the other parameters refer to Table~\ref{tab:params_1}.}
  \label{fig:loss}
\end{figure}

Figure~\ref{fig:loss} shows how the bit-error rate of the preamble decreases during training for the case of the full system with and without restricted energy in comparison to the single agent transmitter from Section~\ref{subsec:single}. The bit-error rate of the preamble represents the loss during training minus the energy penalty. The three depicted runs are based on dissimilar $E_b/N_0$ values during training and necessitate different hyperparameters to converge, thus the absolute BER values should not be compared to each other (We present dependable results about the performance of the resulting modulation schemes in Section~\ref{sec:quant}). However, a qualitative comparison reveals that the single transmitter system converges much faster with less variance because it does not have to come up with a resilient modulation scheme itself but only find the one that the receiver dictates. For the full system, we can see that the variance of the bit error rate is large for the same period of time that the mean energy per symbol (Figure~\ref{fig:unrestricted}) stays relatively high. This is due to the transmitters learning fairly diverse mappings early while exploring the constellation space. Around a certain BER value, which happens to be approximately $10^{-2}$ in this case, the transmitters stick to their discovered schemes and subsequently only decrease the standard deviation of their sampling to lower the BER further. The restricted agent explores faster but longer because of a smaller initial $\text{log}(\sigma)$ which was necessary for convergence.

\subsection{Influence of Hyperparameters and Initialization}

This subsection qualitatively describes the influence of the hyperparameters and the initialization on the learning behavior and convergence of the algorithm. We ran around 4000 hyperparameter sweeps in a random search fashion to find a good set for the evaluation process and to study the influence of each of the parameters on the learning performance. Table~\ref{tab:hyperparams} shows the parameter search space. 

\begin{table}[!h]
\fontsize{9}{9}\selectfont
\renewcommand{\arraystretch}{1.3}
\centering
\begin{tabular}{c c}
\hline
\textbf{Hyperparameter} & \textbf{Range} \\
\hline\hline
Preamble length M: & $128, 256, 512$ \\
Initial log($\sigma$): & $-1.5 ... 1$ \\
Noise power density $N_0$: & $0.01...0.7$ \\
Power loss factor $\lambda_p$: & $0.001 ... 0.1$ \\
Step size: & $0.0001 ... 0.01$ \\k in kNN: & $3$ (hand-picked)\\
Num hidden units: & $40$ (hand-picked)\\
Training iterations: & $2000$ (hand-picked)\\
\end{tabular}
\caption{Value ranges for the hyperparameter sweeps.}
\label{tab:hyperparams}
\end{table}

\subsubsection{Preamble length}

The number of symbols in the preamble determines the sample size for one gradient approximation, since the network parameters are updated after each transmission. While longer preambles offer a more precise reward and thus less noisy update steps, preambles longer than $2^9$ symbols significantly increase training time and become impractical for testing.

\subsubsection{Initial log standard deviation}
The initial log standard deviation of the transmitter plays an important role in the learning process because it controls the level of exploration of the agent in the exploration versus exploitation trade-off. Initial values of the log standard deviation which are too large prevent the agent from learning in a reasonable number of iterations while values that are too small stymie the exploration process and cause the network to prematurely converge on an unstable constellation.

\subsubsection{Step size}
The convergence of the algorithm and the final shape of the constellation is very sensitive to variations in the step size of the gradient update. Small step sizes lead to an excessively long training duration but large step sizes cause an instantaneous splitting of all clusters and low stability. In our experiments we were able to tune a fixed learning rate for the Adam algorithm in a way so that the final constellation automatically converges. However, to guarantee stability across many different conditions it would be wise to implement some kind of learning rate decay.

\subsubsection{Noise power density}
The noise power density in the AWGN channel adds uncertainty to the learning progress, but very low noise levels that do not effect the learning are not realistic. In fact, if the symbol energy is not clipped but only penalized by a soft power loss factor, the agent will simply increase its mean symbol energy to compensate for the noise. In this case, the noise power density is less important to consider (see Section~\ref{sec:quant}).

The noise power density might be an unintuitive measure to the reader, so we provide a conversion chart between values of $N_0$ and the resulting $E_b/N_0$ ratio for standard 16-QAM in Table~\ref{tab:densitytranslations}. As a reference, $E_b/N_0$ = \si{0}{dB} implies that the noise power density in the channel is equal to the mean energy per bit. Every addition of \si{3}{dB} leads to a doubling of the  $E_b/N_0$ ratio. Note that we cannot directly link the noise power density to $E_b/N_0$ values for our learned modulation schemes because the network can vary its mean output energy depending on the current noise level. For a visualization of different noise levels also refer to Figure~\ref{fig:noises}.

\begin{table}[H]
\fontsize{9}{9}\selectfont
\renewcommand{\arraystretch}{1.3}
\centering
\begin{tabular}{c c}
\hline
\textbf{$N_0$} & \textbf{$E_b/N_0$} \\
\hline\hline
0.01 & 11.43dB \\
0.04 & 5.41dB \\
0.09 & 1.88dB \\
0.16 & -0.61dB
\end{tabular}
\caption{Relationship between values of the noise power density $N_0$ used for evaluation and the resulting $E_b/N_0$ values for standard 16-QAM.}
\label{tab:densitytranslations}
\end{table}

\subsubsection{Power constraint}
$\lambda_p$ in Equation~\ref{eq:loss} determines the contribution of the mean output power to the training loss. If the energy per symbol was unconstrained, the constellations points would simply fly off without limit in order to increase the $E_b/N_0$ ratio. However, an excessively high $\lambda_p$ factor inhibits the splitting of the clusters and thus the learning process as a whole. A loss factor for the output power is a soft constraint, because the network can still increase the symbol energy at will if it helps decreasing the BER term. An alternative to the power loss factor is a hard constraint, which forces the output energy of the transmitter to be smaller than or equal to 1. In our approach, we achieve this by normalizing all symbol amplitudes by the value of the largest mean symbol energy of the network if the largest mean energy is greater that 1. We will refer to this as the \enquote{restricted energy} case\footnote{Note that this does not force the energy of each sampled symbols to be strictly smaller than 1, since we only normalize by the mean. The logic behind this is that the final transmitter will no longer use a stochastic policy after learning but only the resulting means of the learning process. Normalizing on a sample basis would stymie the learning progress.}.

\subsubsection{Number of hidden units}
The number of hidden units determines the expressiveness of the neural network which forms the transmitter. However, a more expressive architecture does not necessarily perform better. We found that increasing the number of hidden units past approximately 40 or using even multiple hidden layers does not improve the performance but only slows down the learning.

\subsubsection{k in the kNN receiver}
In the receiver, k determines the number of neighboring points which are considered to determine the bit sequence assigned to a constellation point. As described in Section~\ref{sec:clustering}, it is possible to demodulate a signal with a clustering algorithm very well even for short preamble lengths. However, a powerful demodulator will lead the transmitter to stick with the first best random constellation instead of exploring the space to find a noise-resistant scheme. Hence, we limit k to 3 in order to motivate the transmitter to find a scheme which works even for weak demodulators and is thus more robust. 

\subsubsection{Training iterations}
The number of training iterations determines how often we send the preamble back and forth before we stop the training process and evaluate the resulting modulation scheme. We found that a good set of hyperparameters leads to convergence within the first 1000 to 2000 iterations or less. Hence, we fixed the number of iterations to 2000 since there is no value in evaluating a modulation scheme which has not yet converged. On the other hand, for evaluation we only use the learned constellation points as represented by their means, which converge even faster.

\subsubsection{Initial weights and biases}
\begin{figure}[!ht]
\input{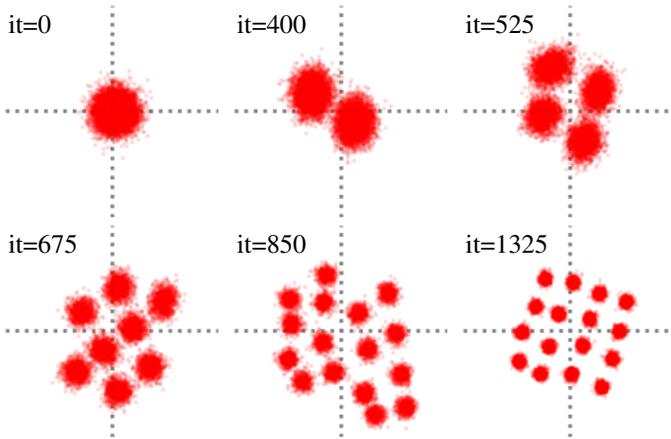}
\caption{Sampled output of the same transmitter as in Figure~\ref{fig:training1} but with a different seed for the random weight generator.}
\label{fig:training2}
\end{figure}

Figure~\ref{fig:training2} shows the sampled output of the same transmitter as in Figure~\ref{fig:training1}, but with another seed for the random generator which provides the initial weights and biases for the network. Although the network converges to an equivalent constellation, the training process runs through very different stages. This effect can also be observed in the high variance of the bit-error rate during some stages of the training (see Figure~\ref{fig:loss}).

We found that initializing the weights and biases of the network with normally distributed values of variance $\sigma=0.2$ and the means with normally distributed values of $\sigma=0.5$ but no bias worked well for most runs. Random initialization proved to be important to break initial symmetries in the network and converge to a meaningful modulation scheme. 
 \newpage
\subsection{Quantitative Analysis} \label{sec:quant}
After describing the training process qualitatively, The following plots show numerical results from the performance analysis of the learned modulation schemes. To generate these plots, we train the system with the provided hyperparameters and then extract the learned modulation scheme represented by the mean output for each possible input bit sequence. We then transmit 10 million symbols modulated with that scheme for different $E_b/N_0$ levels across the channel to the receiver. Since we want to analyze the learned modulation scheme and not the performance of the receiver, we transmit a long preamble of 10,000 symbols to perfectly recreate the constellation points at the receiver by averaging out the noise. Finally, we use the reconstructed mapping to demodulate the noisy samples and measure the bit-error rate of the transmission. 

\begin{figure}[H]
\centering
  \vspace{0.5cm}
\input{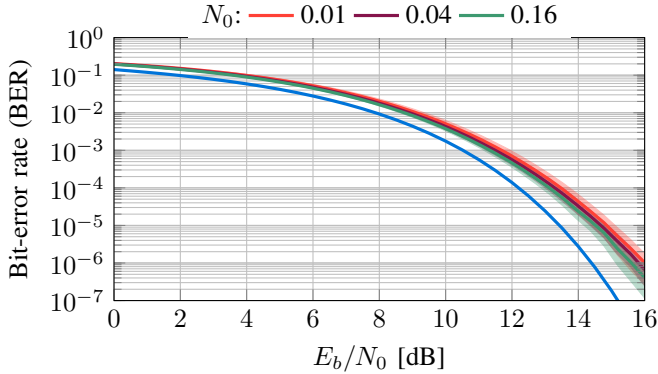}
  \caption{Performance of modulation schemes learned under different noise power densities for the unrestricted energy case. For the other hyperparameters refer to Table~\ref{tab:params_1}. The blue line represents the performance of standard 16-QAM.}
  \label{fig:sweep_noise}
\end{figure}

Figure~\ref{fig:sweep_noise} shows the performance of modulation schemes learned under different noise situations during training with a preamble length of 512 in terms of their BER. We see that although the learned modulation schemes are more error-prone than standard 16-QAM, the learning process is very stable, since varying the noise power density has only a small impact on the result. The high resilience against noise during training is achieved because the network has the option to trade-off symbol energy versus BER in the unrestricted case. In fact, the scheme learned under the very high noise power density of 0.16 performs slightly better than the other ones. This observation can be explained by the fact that the network is forced to learn a very robust transmission scheme, and that the high noise power restricts much of the feasible mapping space as it increases the likelihood for transmission errors. 

\begin{figure}[!ht]
\centering
\input{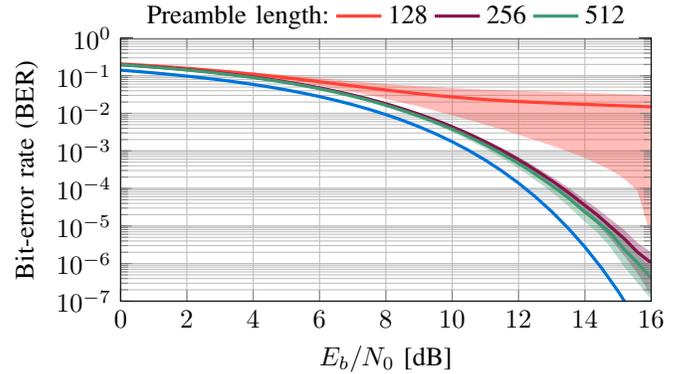}
  \caption{Performance of modulation schemes learned with different preamble lengths under a noise power density of $N_0=0.16$ during training. For the other hyperparameters refer to Table~\ref{tab:params_1}. The blue line represents the performance of standard 16-QAM.}
  \label{fig:sweep_preamble}
\end{figure}

Figure~\ref{fig:sweep_preamble} shows results for a sweep over preamble lengths for a set noise power density. Since the sweeps over a lower $N_0$ were yet again very stable, we were interested to see at which point the learning breaks down and thus chose a very high value of $N_0=0.16$. We see that performance improves for increasing preamble length, which is expected because the longer preamble provides more samples for the score function gradient estimator, improving the robustness of the network. From the graph we can see that the network with preamble length of 128 is extremely variable in performance, as visualized by the high standard deviation for high $E_b/N_0$ ratios. Although there is a significant increase in performance when increasing the preamble length from 128 to 256 symbols, we observe marginal improvement when jumping from 256 to 512 symbols. We predict that further increases to the preamble length would bring performance even close to the 16-QAM baseline, but the trade-off in the number of symbols that need to be transmitted during the learning process would render our approach infeasible for realistic scenarios.

\begin{figure}[!ht]
\centering
\input{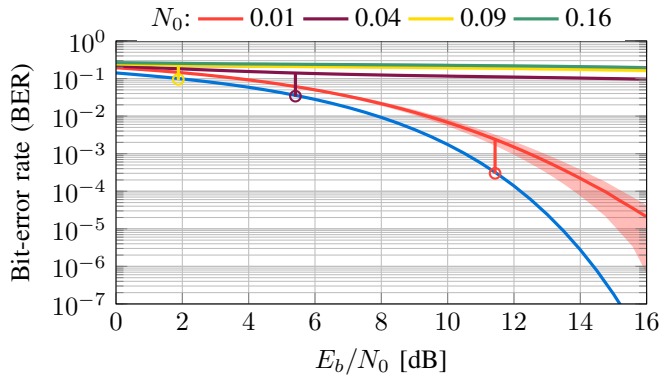}
  \caption{Performance of modulation schemes learned under different noise power densities for the energy restricted case. For the other parameters refer to Table~\ref{tab:params_1}. The blue line represents the performance of standard 16-QAM. The vertical lines refer the noise level during training to the $E_b/N_0$ curve of 16-QAM.}
  \label{fig:sweep_constrained}
\end{figure}

\newpage
Finally, we present results on the performance of modulation schemes learned with restricted symbol energy under different noise levels. In Figure~\ref{fig:training3} we have seen that increasing the noise power density beyond $N_0=0.01$ during training prevents the network from splitting the clusters  all the way to single constellation points. This behavior leads to high bit error rates for these schemes because the constellation points mapped to the same cluster become indistinguishable. Figure~\ref{fig:sweep_constrained} shows that for schemes learned with $N_0\geq0.04$ the BER is high, because the receiver will statistically get 50\% of the bits that differ between constellation points within the same cluster wrong. 

The plot also shows that given a fixed input of four bits per symbol, splitting the constellation into 16 separate points, as represented by the red curve, is always better than keeping them in clusters. The difference between the red curve representing 16 clusters and the purple and yellow curve representing 8 and 4 clusters is staggeringly high for high $E_b/N_0$ values. However, referring the noise level during training back to the performance curve of 16-QAM (vertical lines) reveals that the maximum possible decrease in BER for splitting the clusters is not actually that high for those noise situations. The difference to the feasible red curve is even lower. Thus, we interpret that the network simply has no incentive to split the clusters any further, because that would only marginally decrease the BER for the given noise situation. Even a tenfold increase in learning rate did not change said behavior. While this feature was unanticipated, it is desirable to have the agents automatically switch to a lower order modulation scheme under high noise. However, this assumes that the agents can also exploit those benefits by encoding less bits per symbol, which our architecture currently cannot.

\section{Conclusion and Future Work}\label{sec:conclusion}

The importance of flexible radio design has been confirmed by the initiation of the DARPA Spectrum Collaboration Challenge, which is devoted entirely to the task of building collaborative radio agents via machine learning. In this work, we began by noting the similarities between signal processing in radio domain and machine learning techniques. We tested that observation through a series of preliminary analysis of data-driven approaches to the modulation and the demodulation of digital signals. Using the preliminary results, we formulated the problem of two agents learning modulation schemes in an entirely decentralized fashion and solved it using modern deep reinforcement learning techniques. Finally, we thoroughly analyzed the performance of the algorithm through extensive experiments and discussion. 

We conclude that it \emph{is} possible to learn physical modulation ex-nihilo and decentralized, and that the learning process is very resilient against noise. While it is not surprising that the network tends to learn the standard modulation schemes, these results in conjunction with the highly orchestrated behavior observed during the training process are remarkable. The reward function, which is defined entirely in terms of the bit error rate and the symbol energy, contains no information to induce such structured behavior from the network - there is no reward shaping, no direct measurement of the noise and the neural network itself is rather shallow. At the beginning of training, the reward signal holds almost no useful information because its quality is tied to the performance of the transmitter of the other agent. Nonetheless, the agents learn to balance the exploration versus exploitation trade-off, increase the symbol energy to counteract noise, cluster subsets of constellations points according to their hamming distance, employ local Gray-coding, converge without learning rate decay, implement standard modulation schemes and even try to adapt the modulation scheme to the noise level.

For future work, it would be interesting to increase the capabilities of the agents to deal with more complex settings. A possible next step is to give the agents an option to decide themselves how many bits to encode in each transmitted symbol. Furthermore, the next big step would be to generalize the channel to a non-reciprocal linear time-invariant (LTI) multipath model, which is considerably harder as it is no longer memoryless. Since in this work the BER signal has proven to be rich enough to enable the learning of all kinds of modulation techniques, it is a valid question to ask whether it could also enable the learning of equalization or pre-coding in a decentralized fashion.

% or


\newpage

\appendices
\section{Tables of Hyperparameters}\label{hyps}

\begin{table}[!h]
\fontsize{9}{9}\selectfont
\renewcommand{\arraystretch}{1.3}
\centering
\begin{tabular}{c c}
\hline
\textbf{Hyperparameter} & \textbf{Value} \\
\hline\hline
Preamble length M: & $512$ \\
Initial log($\sigma$): & $-1.0$ \\
Noise power density $N_0$: & $0.01$ \\
Power loss factor $\lambda_p$: & $0.09$ \\
k in kNN: & $3$ \\
Step size: & $0.00245$ \\
Hidden units: & $40$ \\
Training iterations: & $2000$
\end{tabular}
\caption{Hyperparameters if applicable and not stated otherwise.}
\label{tab:params_1}
\end{table}

\begin{table}[!h]
\fontsize{9}{9}\selectfont
\renewcommand{\arraystretch}{1.3}
\centering
\begin{tabular}{c c}
\hline
\textbf{Hyperparameter} & \textbf{Value} \\
\hline\hline
Preamble length M: & $512$ \\
Initial log($\sigma$): & $-2.0$ \\
Noise power density $N_0$: & $0.01$ \\
Power loss factor $\lambda_p$: & $0.05$ \\
k in kNN: & $3$ \\
Step size: & $0.002$ \\
Hidden units: & $40$ \\
Training iterations: & $2000$
\end{tabular}
\caption{Hyperparameters if applicable and not stated otherwise.}
\label{tab:params_2}
\end{table}

%\section*{Acknowledgment}


\ifCLASSOPTIONcaptionsoff
  \newpage
\fi




\bibliographystyle{IEEEtran}

\bibliography{main}
%\begin{thebibliography}{1}


%\end{thebibliography}

%\begin{biography}[{\includegraphics[width=1in,height=1.25in,clip,keepaspectratio]{mshell}}]{Michael Shell}

%\begin{IEEEbiography}[{\includegraphics[width=1in,height=1.25in,clip,keepaspectratio]{picture}}]{John Doe}
%\blindtext
%\end{IEEEbiography}


%\vfill


\end{document}